\documentclass[12pt]{article}
\usepackage{amsfonts}
\usepackage{amsmath}
\usepackage{amssymb}
\usepackage{graphicx}
\usepackage[all, knot]{xy}
\usepackage{tikz}
\usetikzlibrary{cd}

\usepackage{hyperref}

\usepackage[utf8]{inputenc}
\usepackage{epstopdf}
\usepackage[footnotesize]{caption}
\usepackage{amsthm}
\usepackage{enumitem}
\usepackage{mathrsfs}

\usepackage[margin=3cm]{geometry}

\def \be {\begin{equation}}
\def \ee {\end{equation}}
\def \bea {\begin{eqnarray}}
\def \eea {\end{eqnarray}}
\def \nn {\nonumber}

\def \rr {\raise.35ex\hbox{\small $\prime$}\kern-.17em{\mbox{\large $\imath$}}}

\def \dels {\partial\kern-.6em /\kern.1em}
\def \As {{A\kern-.5em / \kern.5em}}
\def \Ds {D\kern-.7em / \kern.5em}

\def \ks {k\kern-.5em /}
\def \ls {l\kern-.5em /}

\def \II {I\hspace{-.1em}I\hspace{.1em}}
\def \III {I\hspace{-.1em}I\hspace{-.15em}I\hspace{.1em}}

\def \sgn {\mbox{\small sgn}}

\newcommand{\ci}[1]{}

\newcommand{\ba}{\begin{eqnarray}}
\newcommand{\ea}{\end{eqnarray}}
\newcommand{\bal}{\begin{align}}
\newcommand{\eal}{\end{align}}
\newcommand{\bay}[1]{\left(\begin{array}{#1}}
\newcommand{\eay}{\end{array}\right)}

\newcommand{\hide}[1]{}

\newlist{axioms}{enumerate}{2}
\setlist[axioms,1]{label=\textbf{A\arabic{axiomsi}.}, ref=A\arabic{axiomsi}}
\setlist[axioms,2]{label=\textbf{A\arabic{axiomsi}\rlap{\myEnumCounter{axiomsii}}.},%
                   ref=A\arabic{axiomsi}\myEnumCounter{axiomsii},%
                   align=parleft,%
                   leftmargin=0em,%
                   itemsep=1.4ex,%
                   before={\stepcounter{axiomsi}}}

  \usetikzlibrary{decorations.markings}

\begin{document}

\begin{titlepage}
\begin{center}

\textbf{\LARGE
Quantum Information in SYK Model
\vskip.3cm
}
\vskip .5in
{\large
Chen-Te Ma$^{a}$ \footnote{e-mail address: yefgst@gmail.com}, 
Jeff Murugan$^{b,c}$ \footnote{e-mail address: jeff.murugan@uct.ac.za}, 
and 
Masaki Tezuka$^d$ \footnote{e-mail address: tezuka@scphys.kyoto-u.ac.jp} 
\\
\vskip 1mm
}
{\sl
$^a$
Centre for Cosmology and Science Popularization, \\
Shree Guru Gobind Singh Tricentenary University, Gurugram, Haryana 122505, India.
\\
$^b$
The Laboratory for Quantum Gravity and Strings,\\
Department of Mathematics and Applied Mathematics,
University of Cape Town, Private Bag, Rondebosch 7700, South Africa.
\\
$^c$
National Institute for Theoretical and Computational Sciences, 
Private Bag X1, Matieland,
South Africa.
\\
$^d$ 
Department of Physics, Kyoto University, Kitashirakawa, Sakyo-ku, Kyoto 606-8502, Japan.
}\\
\vskip 1mm
\vspace{40pt}
\end{center}

\newpage
\begin{abstract} 
\noindent
We investigate the bulk-boundary correspondence in the SYK model from a quantum information perspective. 
The SYK model describes a system of Majorana fermions with random all-to-all interactions, whose disorder average-typically taken over a Gaussian ensemble-admits a dual description in terms of JT gravity in the large-$N$, low-energy limit. 
This framework provides a minimal setting for exploring holography and emergent spacetime in nearly AdS$_2$. 
We probe the holographic principle through diagnostics of quantum chaos and entanglement. 
In the early-time regime, the SYK model saturates the universal bound on the Lyapunov exponent, signaling maximal chaos consistent with semiclassical black hole dynamics. 
In the late-time regime, its spectral statistics are governed by random matrix theory, reflecting universal features of strongly chaotic quantum systems. 
These dynamical properties establish a concrete link between boundary quantum chaos and bulk semiclassical gravity. 
In parallel, we analyze quantum entanglement and the structure of operator algebras to investigate transitions in the associated von Neumann algebras and their implications for emergent geometry. 
To explore the robustness of these phenomena, we consider deformations of the SYK model through modified matter couplings and alternative random distributions. 
Our results clarify how quantum information-theoretic structures encode bulk gravitational dynamics and provide insight into the mechanism of spacetime emergence.
\end{abstract}
\end{titlepage}
\setcounter{tocdepth}{3}
{\hypersetup{linkcolor=black}\tableofcontents}
\newpage
\section{Introduction}
\label{sec:1}
\noindent
Gravitational interaction is the most familiar force in nature, yet its theoretical description remains the most enigmatic.
Unlike the other fundamental forces, which are successfully formulated within quantum field theory (QFT) and made predictive through renormalization, gravity resists a straightforward quantum treatment.
In Quantum Field Theory (QFT), divergences occurring at short distances can be systematically assimilated into a finite array of counterterms.
However, when gravity is treated as a quantum field theory with a dynamical metric, it becomes perturbatively non-renormalizable, requiring an infinite series of counterterms and thereby forfeiting its predictive capacity.
\\

\noindent
At present, the only well-established and experimentally verified description of gravity is Einstein’s theory of general relativity, which governs classical spacetime dynamics.
Beyond the strictly classical paradigm, one may contemplate semiclassical gravity, in which matter fields are quantized on a fixed classical background geometry.
This framework has led to profound insights—such as black hole thermodynamics and Hawking radiation—but it does not provide a fully quantum-mechanical description of spacetime itself.
\\

\noindent
In the pursuit of establishing a coherent quantum theory of gravity, numerous conceptual trajectories have been proposed.
One approach seeks to cure non-renormalizability by introducing non-local interactions with an infinite number of counterterms, or by reformulating gravity as a topological or ultraviolet-complete theory.
An alternative, increasingly influential perspective is that spacetime is not fundamental but emerges.
In this view, a more fundamental quantum theory—possibly formulated without dynamical geometry—gives rise to gravitational dynamics only in appropriate low-energy or collective limits.
In this review, we pursue the latter direction and explore the possibility that gravity emerges from a more fundamental quantum framework.
\\

\noindent
The investigation of Anti–de Sitter (AdS) gravity \cite{Breitenlohner:1982bm} through the framework of conformal field theory (CFT) represents the most concrete realization of the {\it holographic principle} \cite{Maldacena:1997re,Gubser:1998bc,Witten:1998qj}.
The holographic principle posits that the physical degrees of freedom inherent in quantum gravity are {\it wholly} encapsulated within a lower-dimensional boundary theory.
In its most explicit form, the {\it AdS/CFT correspondence} proposes a duality between gravity in $(d+1)$-dimensional AdS spacetime (AdS$_{d+1}$) and a $d$-dimensional conformal field theory (CFT$_d$) defined on its boundary \cite{Maldacena:1997re,Gubser:1998bc,Witten:1998qj}.
\\

\noindent
A prominent example is the duality between type IIB string theory on AdS$_5 \times S^5$ and ${\cal N}=4$ super Yang–Mills (SYM) theory in four dimensions, where ${\cal N}$ denotes the number of supercharges.
In this context, string theory provides a candidate framework for quantum gravity.
In contrast, the dual SYM theory offers a non-gravitational description of the same physics.
This duality implies that a non-perturbative formulation of string theory may be discerned through the lens of gauge theory.
However, in practice, most computations in SU($N$) SYM theory rely on perturbative expansions or semiclassical limits, such as large-$N$ and strong-coupling expansions.
While these methods provide highly nontrivial consistency checks of the correspondence, they do not yet furnish a fully controlled non-perturbative understanding of quantum gravity in generic regimes.
Even in lower-dimensional examples, such as AdS$_3$/CFT$_2$, where pure Einstein gravity is topological and technically simpler, a complete non-perturbative dictionary is known only in special cases.
Understanding the non-perturbative regime is crucial, particularly for addressing fundamental questions such as the dynamics of the metric field and the mechanism by which quantum gravity preserves unitarity and avoids information loss.
Despite significant progress, a comprehensive framework remains elusive.
Motivated by these challenges, we turn to quantum-mechanical models as laboratories for quantum gravity.
Such models often allow for controlled non-perturbative analysis, including numerical simulations, even when analytical methods are limited.
By studying these simplified yet dynamically rich systems, we aim to gain insight into the non-perturbative structure underlying gravitational theories.
\\

\noindent
The endeavor to delineate the gravitational counterpart of a quantum-mechanical model is intricately linked to the boundary characteristics inherent in Jackiw–Teitelboim (JT) gravity \cite{Almheiri:2014cka}.
{\it Pure JT gravity} is a two-dimensional dilaton gravity theory in which the dilaton field couples linearly to the Ricci scalar \cite{Teitelboim:1983ux,Jackiw:1984je}.
Varying the action with respect to the dilaton enforces a constant-curvature constraint, effectively fixing the bulk geometry to AdS$_2$.
Consequently, the bulk theory lacks local propagating gravitational degrees of freedom, and its dynamics are dictated solely by boundary modes. The dimensionless nature of the two-dimensional gravitational coupling, combined with the capability to integrate out the dilaton to enforce the curvature constraint, enables JT gravity to circumvent the conventional perturbative non-renormalizability challenges that beset higher-dimensional gravity.
This renders it a particularly elegant and manageable framework for investigating quantum gravitational fluctuations, particularly those linked to boundary reparameterization modes.
However, despite its simplicity, the boundary description of JT gravity does {\it not} correspond to a conventional conformal field theory.
In particular, the structure of the four-point function—such as the absence of a standard operator-product expansion consistent with a local CFT spectrum—indicates that the dual theory {\it cannot} be a standard CFT$_1$ \cite{Almheiri:2014cka}.
Instead, the boundary dynamics is governed by the {\it Schwarzian theory} \cite{Jensen:2016pah,Maldacena:2016upp,Engelsoy:2016xyb,Bagrets:2016cdf}, which captures reparameterization fluctuations but {\it lacks} the full structure of a conformal field theory.
This subtle boundary characterization presents a central challenge in determining the precise gravitational dual of quantum-mechanical models.
\\

\noindent
A strongly correlated quantum system provides the boundary realization of JT gravity: the Sachdev–Ye–Kitaev (SYK) model \cite{Polchinski:2016xgd,Maldacena:2016hyu}, originally inspired by Kitaev’s proposal \cite{Kitaev:2015}.
The {\it SYK model} describes $N$ Majorana fermions subject to an ensemble of random four-fermion interactions, with stochastic couplings drawn from a Gaussian distribution. 
In the large-$N$ limit, the model becomes solvable after disorder averaging.
In the infrared (IR) regime, its effective action reduces to the {\it Schwarzian theory}. 
\\

\noindent
Under the strict large-$N$ ($N\rightarrow\infty$) and low-energy limits, the SYK two-point and higher-point correlators exhibit emergent conformal symmetry, taking the form of a CFT$_1$. 
The reparameterization symmetry is spontaneously broken down to SL(2, $\mathbb{R}$), generating the Schwarzian action as the effective description of the soft mode. 
In this sense, the SYK model realizes the {\it nearly AdS$_2$/CFT$_1$ correspondence}, which captures the universal {\it low-energy} dynamics of JT gravity.
Historically, the introduction of disorder can be traced back to the Sachdev-Ye model \cite{Sachdev:1992fk}, proposed in the context of strongly correlated electron systems and high-temperature superconductivity.
Kitaev later refined the Hamiltonian structure to reveal its holographic interpretation, where the model becomes {\it maximally chaotic} and non-integrable.
\\

\noindent
Quantum chaos is itself a profound and subtle subject.
In classical mechanics, chaos \cite{Guckenheimer:1979,Milnor:1985,Banks:1992} is characterized by profound sensitivity to initial conditions: trajectories within phase space diverge exponentially over time.
However, in quantum mechanics, the notion of a trajectory is absent, since time evolution is governed by the Schrödinger equation and constrained by the uncertainty principle \cite{Berry:1977zz}.
Consequently, the traditional diagnostics of chaos are {\it not} directly applicable to quantum systems.
\\

\noindent
Even in the semiclassical limit $\hbar \to 0$, the relationship between classical and quantum chaos is nontrivial.
The limits of vanishing $\hbar$ and vanishing non-integrability parameters (which quantify deviation from integrability) do {\it not} generally commute.
Consequently, taking the semiclassical limit does not smoothly reproduce classical chaotic dynamics \cite{Berry:1977zz}.
This non-commutativity reflects the intrinsic structural differences between quantum and classical evolution.
\\

\noindent
Chaos in quantum systems is diagnosed through out-of-time-order correlators (OTOCs), constructed from Heisenberg-evolved operators \cite{Larkin:1969,Maldacena:2015waa,Narovlansky:2025tpb},
\bea
{\cal O}(t)\equiv e^{i H t}{\cal O}(0) e^{-iH t},
\eea
The growth of the squared commutator (or anti-commutator for fermions) measures {\it sensitivity} to initial conditions \cite{Maldacena:2015waa}.
In chaotic systems, the OTOC exhibits {\it exponential growth} with Lyapunov exponent $\lambda_L$.
The SYK model saturates the universal {\it chaos bound} $\lambda_L \le 2\pi/\beta$ \cite{Maldacena:2015waa}, where $\beta$ is the inverse temperature, signaling an emergent gravitational dual described by JT gravity.
More generally, in two- or higher-dimensional CFTs, saturation of the chaos bound is associated with {\it Einstein gravity} in the bulk \cite{Maldacena:2015waa,Narovlansky:2025tpb}.
Furthermore, the spectral statistics of the SYK model exhibit random matrix theory (RMT) \cite{Dyson:1962es,You:2016ldz} behavior, including level repulsion \cite{Bohigas:1983er} and the spectral form factor \cite{Brezin:1997rze} with the ramp-plateau structure \cite{Garcia-Garcia:2016mno,Garcia-Garcia:2017pzl,Dyer:2016pou,Cotler:2016fpe,Krishnan:2016bvg}, as expected for quantum chaotic systems.
Therefore, the SYK model {\it not only} provides a concrete realization of the nearly AdS$_2$/CFT$_1$ correspondence but also establishes a deep connection among quantum chaos, random matrix universality, and emergent gravitational dynamics.
\\

\noindent
From a statistical perspective, quantum chaos is assessed through spectral statistics.
According to the Bohigas–Giannoni–Schmit (BGS) conjecture \cite{Bohigas:1983er}, the energy-level statistics of quantum systems \cite{Oganesyan:2007wpd,Atas:2013gvn,Nishigaki:2024yjr} whose classical counterparts are chaotic follow the predictions of RMT. 
In contrast, integrable systems exhibit {\it Poisson} level statistics \cite{Berry:1977wpp}.
This hypothesis was subsequently substantiated through semiclassical analysis, in which a correspondence between {\it quantum spectra} and {\it classical periodic orbits} was established using Gutzwiller’s trace formula \cite{Muller:2004nb}.
These developments provide a deep bridge between quantum and classical descriptions.
However, the traditional framework of quantum chaos assumes systems {\it without} disorder.
When disorder is introduced, as in models {\it with} random couplings, the distinction between integrable and non-integrable configurations becomes more subtle.
Disorder averaging possibly {\it mixes} integrable and chaotic realizations, and apparent chaotic behavior can emerge from statistical randomness itself \cite{Lau:2018kpa,Lau:2020qnl,Ozaki:2025mma}.
For example, certain variants of the SYK model with specific coupling distributions can admit integrable structures \cite{Ozaki:2025mma,Fukai:2025dcq}.
Nevertheless, after disorder averaging, the spectral statistics typically follow RMT predictions, and the model exhibits signatures of quantum chaos.
\\

\noindent
In disordered systems, conventional diagnostics of integrability—such as the conserved charges—become less transparent.
As a result, the definition of quantum chaos in the presence of disorder is more subtle and still partly guided by analogies to clean systems.
Various SYK-type variants have been meticulously constructed to elucidate this matter and to investigate the intricate interplay between disorder, holography, and quantum chaos, as well as the categorization of chaotic and integrable phases \cite{Moitra:2022glw,Lau:2023pot,Lau:2025dgd}.
\\

\noindent
An even more fundamental challenge arises in few-body quantum systems.
Most diagnostic tools—such as spectral statistics and Lyapunov exponents—are naturally formulated in many-body systems with {\it large} Hilbert spaces.
Spectral statistical methods are meaningful primarily in the limit of large degrees of freedom.
For finite systems, the Lyapunov exponent may depend {\it sensitively} on the choice of operator basis or Hilbert space embedding. While the concept of integrability remains well-defined for few-body Hamiltonians without disorder, a universally accepted diagnostic of few-body quantum chaos is still lacking.
Therefore, developing new tools and conceptual frameworks for diagnosing quantum chaos—especially in few-body systems and in the presence of disorder—remains one of the most important open problems in the field.
\\

\noindent
Although deforming the SYK model to engineer a controlled JT gravity dual is technically straightforward \cite{Davison:2016ngz,Anninos:2020cwo,Anninos:2022qgy}, the deeper motivation for studying holography goes beyond reproducing nearly AdS$_2$/CFT$_1$.
A central question is how spacetime itself emerges from a more fundamental quantum description.
One powerful framework for addressing this issue is the classification of operator algebras in terms of von Neumann (vN) algebras \cite{Leutheusser:2021qhd,Chandrasekaran:2022cip}.
\\

\noindent
Von Neumann algebras are categorized into three principal types.
Type I algebras correspond to ordinary quantum-mechanical systems with a well-defined Hilbert space factorization and density matrices.
They contain all bounded operators on a Hilbert space and admit pure states.
A complete, non-perturbative boundary theory—such as the microscopic description of quantum gravity—is expected to correspond to a Type I algebra \cite{Leutheusser:2021qhd,Chandrasekaran:2022cip}.
Type \II algebras do {\it not} admit minimal projectors and therefore lack pure states in the usual sense, though they still possess a trace.
These algebras naturally arise in semiclassical gravitational systems, where one can define generalized entropies but not a conventional Hilbert space factorization \cite{Leutheusser:2021qhd,Chandrasekaran:2022cip}.
Type \III algebras, which appear in local quantum field theory, {\it lack} a trace and do {\it not} admit a density matrix description for the spatial subregions.
In such systems, the concept of entropy eludes definition without regularization.
Pure gravity or boundary theories at leading order in $1/N$ restricted to low-energy bulk sectors are often associated with Type \III structures \cite{Leutheusser:2021qhd,Chandrasekaran:2022cip}.
The distinction between these algebraic types has been proposed as a precise way to characterize the emergence of spacetime and the transition between microscopic and semiclassical descriptions \cite{Leutheusser:2021qhd,Chandrasekaran:2022cip}.
However, the vN algebra framework assumes that a theory can be classified purely by its bounded operator algebra.
To avoid ambiguities associated with starting from an already coarse-grained algebra (such as Type \III), it is conceptually cleaner to begin with a well-defined Type I system—such as the SYK model—and study how lower-type structures emerge dynamically \cite{Chandrasekaran:2022qmq}.
\\

\noindent
One practical realization is to deform the SYK model by coupling source terms to boundary operators, in accordance with the AdS/CFT dictionary.
The source corresponds to the boundary value of bulk matter fields \cite{Moitra:2022glw}.
This procedure yields a consistent boundary effective action dual to JT gravity with matter.
Because the boundary Hilbert space now factorizes into matter and gravitational (reparameterization) sectors, one can compute entanglement measures between these sectors and investigate the emergent algebraic structure of the graviton degrees of freedom \cite{Lau:2023pot,Lau:2025dgd}.
For bosonic matter fields, deformations typically require a UV cutoff for regularization.
In contrast, fermionic matter fields provide a technically cleaner setup.
By deforming the Gaussian distribution of random couplings, one can go beyond purely quadratic matter theories and generate controlled interactions \cite{Lau:2023pot,Lau:2025dgd}.
\\

\noindent
Importantly, in the strict large-$N$ limit, the theory reduces to JT gravity.
The corrections in $1/N$ introduce nonlocal gravitational effects in the bulk \cite{Gross:2017hcz}.
If new perturbation parameters are introduced independently of the $1/N$ expansion, the resulting framework becomes more general than a standard perturbative bulk expansion, providing a broader class of holographic correspondences.
Therefore, deformations of the SYK model furnish a clean and controllable laboratory for exploring foundational questions of quantum gravity—particularly the emergence of spacetime, the algebraic structure of gravitational degrees of freedom, and the transition from microscopic Type I descriptions to Type \II/\III regimes.
These insights are difficult to access directly in higher-dimensional gravitational theories.

\subsection{Outline}
\noindent
The outline of this review is as follows.
In Sec.~\ref{sec:2} and Sec.~\ref{sec:3}, we introduce the necessary background on Jackiw–Teitelboim (JT) gravity and the Sachdev–Ye–Kitaev (SYK) model.
These sections review the boundary description of pure JT gravity and explain how the low-energy effective theory of the SYK model reproduces the Schwarzian dynamics associated with nearly AdS$_2$/CFT$_1$.
\\

\noindent
In Sec.~\ref{sec:4}, we discuss quantum chaos, beginning with classical chaos.
We first review the classical definition in terms of sensitivity to initial conditions, focusing on bounded phase-space intervals.
We then extend the discussion to quantum systems, introducing modern diagnostics of quantum chaos such as out-of-time-order correlators (OTOCs) and spectral statistics.
\\

\noindent
Section~\ref{sec:5} presents deformations of the SYK model designed to generate matter couplings in the large-$N$ limit.
We begin by reviewing the boundary description of JT gravity and the AdS/CFT dictionary.
We then extend these ideas to construct controlled deformations of the SYK model and analyze their implications for quantum chaos and emergent gravitational dynamics.
In Sec.~\ref{sec:6}, we review the framework of von Neumann (vN) algebras and discuss how deformed SYK models provide a useful laboratory for studying the emergence of spacetime from an algebraic perspective.
We present the mathematical structure of vN-algebra classifications and connect them to the physical interpretation of the gravitational and matter sectors in the boundary theory.
Finally, Sec.~\ref{sec:7} summarizes the main results and outlines future directions and open problems related to quantum chaos, holography, SYK deformations, and the algebraic structure of emergent gravity.

\section{JT Gravity}
\label{sec:2}
\noindent
We introduce the necessary background to JT gravity \cite{Teitelboim:1983ux,Jackiw:1984je}.
Beginning from the Lagrangian formalism, we review the equations of motion to analyze the classical solution space \cite{Teitelboim:1983ux,Jackiw:1984je}.
We then derive the boundary description using Schwinger-Keldysh theory \cite{Jensen:2016pah,Maldacena:2016upp}.

\subsection{Lagrangian}
\noindent
We first introduce the Lagrangian of pure JT gravity.
The JT gravity action, $S_{\mathrm{JT}}$, consists of a bulk and a boundary contribution \cite{Teitelboim:1983ux,Jackiw:1984je},
\bea
&&
S_{\mathrm{JT}}
\nn\\
&=&
-\frac{1}{16\pi G_2}\int_{\mathcal{M}} d^2x\, \sqrt{|\det g_{\mu\nu}|}\, \phi (R - 2\Lambda)
- \frac{1}{8\pi G_2}\int_{\partial \mathcal{M}} du\, \sqrt{|\det h_{uu}|}\,\phi K,
\eea
where the $G_2$ is the 2D gravitational constant.
The dilaton field $\phi$ couples linearly to the Ricci scalar
\bea
R\equiv g^{\mu\nu}R_{\mu\nu},
\eea
where
\bea
R_{\mu\nu}&\equiv&\partial_{\delta}\Gamma^{\delta}_{\nu\mu}-\partial_{\nu}\Gamma^{\delta}_{\delta\mu}
+\Gamma^{\delta}_{\delta\lambda}\Gamma^{\lambda}_{\nu\mu}
-\Gamma^{\delta}_{\nu\lambda}\Gamma^{\lambda}_{\delta\mu}; \
\Gamma^{\mu}_{\nu\delta}\equiv\frac{1}{2}g^{\mu\lambda}\bigg(\partial_{\delta}g_{\lambda\nu}+\partial_{\nu}g_{\lambda\delta}
-\partial_{\lambda}g_{\nu\delta}\bigg).
\nn\\
\eea
enforcing constant curvature via its equations of motion.
The boundary term ensures a well-posed variational principle under Dirichlet boundary conditions, which fix the boundary data on $\partial \mathcal{M}$,
\bea
\delta g_{\mu\nu}=0; \ \delta\phi=0; \ \frac{\partial\delta g_{\mu\nu}}{\partial u}=0.
\eea
Here, $h_{uu}$ denotes the induced boundary metric, while the extrinsic curvature is given by,
\bea
K \equiv g^{\mu\nu}\nabla_{\nu} n_{\mu},,
\eea
where $\nabla_{\nu}$ is a convariant derivative, and $n^{\mu}$ is a unit normal vector satisfying,
\bea
n^{\rho}n_{\rho}=\left\{\begin{array}{ll}
-1, & \mbox{if $\Lambda> 0$}; \\
1, & \mbox{if $\Lambda<0$}.
\end{array} \right.
\eea
For the AdS$_2$ case, the cosmological constant is negative, $\Lambda < 0$.

\subsection{Equations of Motion}
We now present the equations of motion for the dilaton and metric fields \cite{Teitelboim:1983ux,Jackiw:1984je}.
The variation of the dilaton field determines the geometry through its linear coupling to the scalar curvature.
Varying the metric field gives the equation the dilaton follows.

\subsubsection{Dilaton Field}
\noindent
We first vary the dilaton field.
The equation of motion is \cite{Teitelboim:1983ux,Jackiw:1984je}
\bea
R=2\Lambda.
\eea
When we consider a negative cosmological constant, the solution shows the unique AdS$_2$ \cite{Teitelboim:1983ux,Jackiw:1984je}
\bea
ds^2=-\frac{1}{\Lambda}\frac{dt^2+dz^2}{z^2}.
\eea
Consequently, the geometry remains immutable.

\subsubsection{Metric Field}
\noindent
The variation of the metric shows the equations of motion \cite{Teitelboim:1983ux,Jackiw:1984je}
\bea
\nabla_{\mu}\nabla_{\nu}\phi
-g_{\mu\nu}\nabla^2\phi-\Lambda g_{\mu\nu}\phi=0.
\eea
Here, we use the following fact
\bea
R_{\mu\nu}-\frac{1}{2}Rg_{\mu\nu}=0.
\eea
Hence, the solution of the dilaton field follows the equation of motion for the metric field.

\subsection{Conformal Gauge}
\noindent
We choose the conformal gauge to rewrite the metric:
\bea
ds^2=-2e^{2\rho}dx^+dx^-; \ e^{2\rho}=-\frac{2}{\Lambda}\frac{1}{(x^+-x^-)^2},
\eea
where
\bea
x^+\equiv t+iz; \ x^-\equiv t-iz.
\eea
The equations of motion may be reformulated as follows:
\bea
-\nabla_+\nabla_+\phi&=&-\partial_+^2\phi+\Gamma^+_{++}\partial_+\phi=-\partial_+^2\phi+2\partial_+\rho\partial_+\phi,
\nn\\
-\nabla_-\nabla_-\phi&=&-\partial_-^2\phi+\Gamma^-_{--}\partial_-\phi=-\partial_-^2\phi+2\partial_-\rho\partial_-\phi,
\nn\\
-\nabla_+\nabla_-\phi&=&-\partial_+\partial_-\phi,
\nn\\
\nabla^2\phi&=&-2 e^{-2\rho}\partial_+\partial_-\phi.
\eea
Hence, the equations of motion become:
\bea
\partial_{\pm}(e^{-2\rho}\partial_{\pm}\phi)=0; \ \partial_+\partial_-\phi-\Lambda e^{2\rho}\phi=0.
\eea
Using the conformal gauge makes it easier to obtain the dilaton solution
\bea
\phi=\frac{a+b(x^++x^-)+c(x^+x^-)}{x^+-x^-},
\label{solbulk}
\eea
where $a, b, c$ are arbitrary constants.
We will use it to examine the solution space for the boundary description of the JT gravity \cite{Maldacena:2016upp}.

\subsection{Boundary Description}
\noindent
For the case of a non-vanishing metric, the induced metric on the boundary is
\bea
ds^2_{b}\equiv h_{uu}du^2=-\frac{1}{\Lambda}\frac{t^{\prime 2}+z^{\prime 2}}{z^2}du^2,
\eea
where
\bea
t^{\prime}\equiv \frac{dt}{du}; \ z^{\prime}\equiv\frac{dz}{du}.
\eea
We fix the proper length of the boundary curve \cite{Jensen:2016pah,Maldacena:2016upp}:
\bea
\frac{1}{|\Lambda|}\frac{t^{\prime 2}+z^{\prime 2}}{z^2}=\frac{1}{\epsilon^2}.
\eea
If we assume that $z^{\prime 2}\ll t^{\prime 2}$, we can obtain \cite{Jensen:2016pah,Maldacena:2016upp}
\bea
t^{\prime 2}=\frac{|\Lambda|}{\epsilon^2}z^2+\cdots
\eea
or
\bea
z^2=\frac{\epsilon^2}{|\Lambda|}t^{\prime 2}+\cdots.
\label{smallz}
\eea
We can choose the tangent curve $(t^{\prime}, z^{\prime})$ \cite{Jensen:2016pah,Maldacena:2016upp}.
Because the tangent curve is orthogonal to a unit normal vector, the unit normal vector is \cite{Jensen:2016pah,Maldacena:2016upp}
\bea
n^t=\sqrt{|\Lambda|}\frac{zz^{\prime}}{\sqrt{t^{\prime 2}+z^{\prime 2}}}; \ n^z=-\sqrt{|\Lambda|}\frac{zt^{\prime}}{\sqrt{t^{\prime 2}+z^{\prime 2}}}.
\eea
The normalization condition is \cite{Jensen:2016pah,Maldacena:2016upp}
\bea
g_{\mu\nu}n^{\mu}n^{\nu}=1.
\eea
The trace of the extrinsic curvature is \cite{Jensen:2016pah,Maldacena:2016upp}:
\bea
K&=&g^{\mu\nu}\nabla_{\nu}n_{\mu}=g^{\mu\nu}\big(\partial_{\nu}n_{\mu}-\Gamma^{\gamma}_{\mu\nu}n_{\gamma}\big)
\nn\\
&=&\sqrt{|\Lambda|}\frac{t^{\prime}}{\sqrt{t^{\prime 2}+z^{\prime 2}}}
-\sqrt{|\Lambda|}\frac{zz^{\prime}t^{\prime\prime}}{(t^{\prime 2}+z^{\prime 2})^{\frac{3}{2}}}
+\sqrt{|\Lambda|}\frac{zt^{\prime}z^{\prime\prime}}{(t^{\prime 2}+z^{\prime 2})^{\frac{3}{2}}}.
\eea
Now we do expansion up to the order of $\epsilon^2$ \cite{Jensen:2016pah,Maldacena:2016upp}:
\bea
K=\sqrt{|\Lambda|}+\frac{\epsilon^2}{\sqrt{|\Lambda|}}\bigg(\frac{t^{\prime\prime\prime}}{t^{\prime}}-\frac{3}{2}\frac{t^{\prime\prime 2}}{t^{\prime 2}}\bigg)+\cdots.
\eea
Therefore, we can find that the trace of the extrinsic curvature has the Schwarzian term \cite{Jensen:2016pah,Maldacena:2016upp}
\bea
Sch(t, u)\equiv\frac{t^{\prime\prime\prime}}{t^{\prime}}-\frac{3}{2}\frac{t^{\prime\prime 2}}{t^{\prime 2}}.
\eea
We choose the dilaton field on the boundary as in \cite{Jensen:2016pah,Maldacena:2016upp}
\bea
\phi\rightarrow\frac{\sqrt{|\Lambda|}}{\epsilon}\phi_b.
\eea
Hence, we obtain the Schwarzian theory \cite{Jensen:2016pah,Maldacena:2016upp}
\bea
-\frac{1}{8\pi G_2}\int du\sqrt{|h_{uu}|}\ \phi K
=-\frac{1}{8\pi G_2}\int du\ \phi_b\ Sch(t, u)
-\frac{1}{8\pi G_2}\int du\ \frac{\sqrt{|\Lambda|}}{\epsilon^2}\phi_b.
\eea
Now, $t$ is a dynamical field instead of a time variable \cite{Maldacena:2016upp}.
\\

\noindent
Now, we take the variation with respect to $t$ and show the equation of motion:
\bea
&&
Sch(t, u)=\frac{t^{\prime\prime\prime}}{t^{\prime}}-\frac{3}{2}\frac{t^{\prime\prime 2}}{t^{\prime 2}}= -\frac{1}{2}W^2+W^{\prime}; \ W\equiv\frac{t^{\prime\prime}}{t^{\prime}},
\nn\\
&&
\bigg\lbrack\bigg(\phi_b W+\phi_b^{\prime}\bigg)^{\prime}\frac{1}{t^{\prime}}\bigg\rbrack^{\prime}
=\bigg\lbrack\frac{1}{t^{\prime}}\bigg(\frac{\big(t^{\prime}\phi_b\big)^{\prime}}{t^{\prime}}\bigg)^{\prime}\bigg\rbrack^{\prime}=0.
\eea
The general solution of $\phi_b$ is
\bea
\phi_b=\frac{\tilde{\alpha}+\tilde{\beta}t+\tilde{\gamma}t^2}{t^{\prime}}.
\eea
When we consider $z\rightarrow 0$ as in Eq. \eqref{smallz}, the solution of Eq. \eqref{solbulk} is the general solution of $\phi_b$ \cite{Maldacena:2016upp}.
Hence, we can treat $t$ as a dynamical field, with a consistent solution space linking JT gravity and the boundary description, Schwarzian theory \cite{Maldacena:2016upp}.

\section{SYK Model}
\label{sec:3}
\noindent
We begin from the Hamiltonian formalism of the SYK model and then compute the partition function \cite{Polchinski:2016xgd,Maldacena:2016hyu}.
From the partition function, we can obtain the Schwinger-Dyson (SD) equation to extract the Green's function.
Finally, we use the Green's function result to obtain the Schwarzian theory.

\subsection{Hamiltonian}
\noindent
While the SYK model has been generalized to $q$-body interactions \cite{Maldacena:2016hyu} and complex fermions \cite{Davison:2016ngz}, here we introduce the model in its original form, involving $q=4$ interactions between Majorana fermions.
The SYK model is articulated through the Hamiltonian,
\begin{eqnarray}
H_{\mathrm{SYK}} = \sum_{1 \le i_1 < i_2 < i_3 < i_4 \le N} J_{i_1 i_2 i_3 i_4} \psi_{i_1} \psi_{i_2} \psi_{i_3} \psi_{i_4},
\end{eqnarray}
where $\{\psi_i\}_{i=1}^N$ represents Majorana fermion operators that adhere to the anticommutation relations $\{\psi_i, \psi_j\} = \delta_{ij}$, and $J_{i_1 i_2 i_3 i_4}$ denotes random coupling constants that are independently sampled from a Gaussian distribution characterized by a zero mean and variance,
\begin{eqnarray}\label{GD}
\overline{J_{i_1 i_2 i_3 i_4}^2} = \frac{6J^2}{N^{3}},
\end{eqnarray}
where $J$ is a constant that establishes the energy scale of the model.
In the large-$N$ limit, it is particularly convenient to analyze the bulk gravitational theory through the SYK model.
The Schwarzian action governs the low-energy limit of the SYK model.
This correspondence is valid in the regime
\bea
G_2 \sim \frac{1}{N} \ll 1; \ \phi_b \sim \frac{1}{\beta J} \ll 1,
\eea
where $\beta$ is the inverse temperature.

\subsection{Partition Function}
\noindent
The partition function of the SYK model is \cite{Polchinski:2016xgd,Maldacena:2016hyu}
\bea
&&
Z_{\mathrm{SYK}}
\nn\\
&=&\int dj\ \int D\psi\ \exp\Bigg\lbrack-\int d\tau\Bigg(\frac{1}{2}\sum_{j=1}^N\psi_j\dot{\psi}_j
+\sum_{1<i_1<i_2<i_{3}<i_4\le N}j_{i_1i_2i_{3}i_4}\psi_{i_1}\psi_{i_2}\psi_{i_{3}}\psi_{i_4}\Bigg)\Bigg\rbrack
\nn\\
&\times&\exp\Bigg(-\sum_{1<i_1<i_2<i_{3}<i_4\le N}j_{i_1i_2i_{3}i_4}^2\frac{N^{3}}{12J^2}\Bigg),
\eea
where $\tau$ is a time coordinate.
We can integrate the random coupling constant out to get the partition function of the SYK model,
\bea
&&
Z_{\mathrm{SYK}}
\nn\\
&=&\int D\psi\ \exp\Bigg\lbrack-\frac{1}{2}\int d\tau\ \sum_{j=1}^N\psi_j(\tau)\dot{\psi}_j(\tau)
+\frac{J^2}{8}N\int d\tau d\tilde{\tau}\ \Bigg(\frac{1}{N}\sum_{l=1}^N\psi_l(\tau)\psi_l(\tilde{\tau})\Bigg)^4\Bigg\rbrack,
\nn\\
\eea
in which the integration of the random coupling constant is equivalent to using
\bea
j_{i_1 i_2i_{3}i_4}=\frac{6J^2}{N^{3}}\int d\tau\ \psi_{i_1}\psi_{i_2}\psi_{i_{3}}\psi_{i_4}.
\eea
Now we insert
\bea
1&=&\int D\tilde{G}\ \delta\bigg(\tilde{G}(\tau, \tilde{\tau})-\frac{1}{N}\sum_{j=1}^N\psi_j(\tau)\psi_j(\tilde{\tau})\bigg)
\nn\\
&\sim&\int D\tilde{G}D\tilde{\Sigma}\ \exp\Bigg\lbrack-\int d\tau d\tilde{\tau}\ \frac{\tilde{\Sigma}(\tau, \tilde{\tau})}{2}\Bigg(N\tilde{G}(\tau, \tilde{\tau})-\sum_{l=1}^N\psi_l(\tau)\psi_l(\tilde{\tau})\Bigg)\Bigg\rbrack
\eea
into the partition function above:
\bea
&&
Z_{\mathrm{SYK}}
\nn\\
&\sim&\int D\psi D\tilde{G} D\tilde{\Sigma}\  \exp\Bigg\lbrack-\frac{1}{2}\int d\tau\ \sum_{j=1}^N\psi_j(\tau)\dot{\psi}_j(\tau)
+\frac{J^2}{8}N\int d\tau d\tilde{\tau}\ \Bigg(\frac{1}{N}\sum_{l=1}^N\psi_l(\tau)\psi_l(\tilde{\tau})\Bigg)^4\Bigg\rbrack
\nn\\
&&\times\exp\Bigg\lbrack-\int d\tau d\tilde{\tau}\ \frac{\tilde{\Sigma}(\tau, \tilde{\tau})}{2}\Bigg(N\tilde{G}(\tau, \tilde{\tau})-\sum_{l=1}^N\psi_l(\tau)\psi_l(\tilde{\tau})\Bigg)\Bigg\rbrack.
\eea
Then we integrate the Majorana fermion fields ($\psi$) out to obtain
\bea
&&
Z_{\mathrm{SYK}}
\nn\\
&\sim&\int D\tilde{G} D\tilde{\Sigma}\ \exp\bigg\{ N\bigg\lbrack\ln\mbox{Pf}\bigg(\partial_{\tau}-\tilde{\Sigma}\bigg)
-\frac{1}{2}\int d\tau d\tilde{\tau}\ \bigg(\tilde{G}(\tau, \tilde{\tau})\tilde{\Sigma}(\tau, \tau^{\prime})-\frac{J^2}{4}\tilde{G}^4(\tau, \tilde{\tau})\bigg)\bigg\rbrack\bigg\}
\nn\\
&=&\int D\tilde{G} D\tilde{\Sigma}\ \exp\bigg\{ N\bigg\lbrack\frac{1}{2}\ln\det\bigg(\partial_{\tau}-\tilde{\Sigma}\bigg)
-\frac{1}{2}\int d\tau d\tilde{\tau}\ \bigg(\tilde{G}(\tau, \tilde{\tau})\tilde{\Sigma}(\tau, \tilde{\tau})-\frac{J^2}{4}\tilde{G}^4(\tau, \tilde{\tau})\bigg)\bigg\rbrack\bigg\}
\nn\\
&\equiv&\int D\tilde{G}D\tilde{\Sigma}\ e^{-S_{\mathrm{SYK}}},
\eea
where $S_{\mathrm{SYK}}$ is an effective action of the SYK model
\bea
\frac{S_{\mathrm{SYK}}}{N}=-\frac{1}{2}\ln\det\bigg(\partial_{\tau}-\tilde{\Sigma}\bigg)+\frac{1}{2}\int d\tau d\tilde{\tau}\ \bigg(\tilde{\Sigma}(\tau, \tilde{\tau})\tilde{G}(\tau, \tilde{\tau})-\frac{J^2}{4}\tilde{G}^4(\tau, \tilde{\tau})\bigg).
\eea

\subsection{Schwinger-Dyson Equation}
\noindent
We can take the variation of the effective action of the SYK model with respect to $\tilde{\Sigma}$ to obtain one SD equation \cite{Polchinski:2016xgd,Maldacena:2016hyu}
\bea
\bigg(\delta(\tau, \tilde{\tau})\partial_{\tau}-\tilde{\Sigma}(\tau, \tilde{\tau})\bigg)^{-1}=\tilde{G}(\tau, \tilde{\tau}).
\label{eqn:SD1}
\eea
We then take a variation of the effective action with respect to $\tilde{G}$ to obtain another SD equation
\bea
\tilde{\Sigma}(\tau, \tau^{\prime})=J^2\tilde{G}^{3}(\tau, \tilde{\tau}).
\label{eqn:J2G3}
\eea
The above SD equations are determined from the saddle point
of the effective action of the SYK model.
If we consider the low-temperature limit ($\beta J\gg 1$), we can ignore the time-derivative term in Eq. \eqref{eqn:SD1} and obtain:
\bea
\label{sds}
\int d\tilde{\tau}\ \tilde{G}(\tau, \tilde{\tau})\tilde{\Sigma}(\tilde{\tau}, \bar{\tau})
=-\delta(\tau-\bar{\tau}); \
\tilde{\Sigma}(\tau,\tilde{\tau})=J^2\big(\tilde{G}(\tau,\tilde{\tau})\big)^{3}.
\eea
The SD equations at the low-temperature limit are invariant under the following reparametrization symmetry \cite{Maldacena:2016hyu}:
\bea
\tilde{G}(\tau, \tilde{\tau})\rightarrow\big(f^{\prime}(\tau)f^{\prime}(\tilde{\tau})\big)^{\frac{1}{4}}\tilde{G}\big(f(\tau), f(\tilde{\tau})\big); \
\tilde{\Sigma}(\tau, \tilde{\tau})\rightarrow\big(f^{\prime}(\tau)f^{\prime}(\tilde{\tau})\big)^{\frac{3}{4}}\tilde{\Sigma}\big(f(\tau), f(\tilde{\tau})\big).
\eea

\subsection{Green's Function}
\noindent
We could use the conformal ansatz of the Green's function to solve the SD equations in the low-energy limit \cite{Polchinski:2016xgd,Maldacena:2016hyu},
\bea
\tilde{G}(\tau)=\frac{b}{|\tau|^{2\Delta}}\sgn(\tau); \
\tilde{\Sigma}(\tau)=J^2\tilde{G}^{3}(\tau),
\eea
where $b$ is a constant.
The sign function is defined as follows
\bea
\sgn(x)=\left\{\begin{array}{ll}
1, & \mbox{if $x> 0$}; \\
0, & \mbox{if $x=0$}; \\
-1, & \mbox{if $x<0$}.
\end{array} \right.
\eea
We choose the conformal dimension as
\bea
\Delta=\frac{1}{4}
\eea
because the SYK model in the IR limit is invariant under the scaling:
\bea
\tau\rightarrow\lambda\tau; \ \psi_j\rightarrow \lambda^{-\Delta}\psi_j.
\eea
We can determine the Green's function in momentum space through the following integration results:
\bea
\int_{-\infty}^{\infty}d\tau\ e^{i\omega\tau}\ \frac{\sgn(\tau)}{|\tau|^{2\Delta}}
&=&2i\mathrm{Im}\bigg(\int_0^{\infty}d\tau\ e^{i\omega\tau}\frac{1}{\tau^{2\Delta}}\bigg)
=2i\mathrm{Im}\bigg\lbrack\bigg(\frac{i}{\omega}\bigg)^{1-2\Delta}\Gamma(1-2\Delta)\bigg\rbrack
\nn\\
&=&
2i\cos(\pi\Delta)\Gamma(1-2\Delta)\frac{1}{\omega^{1-2\Delta}}.
\eea
By using the gamma function in the integration formula:
\bea
\Gamma(z)\equiv\int_0^{\infty}dt\ t^{z-1}e^{-t},
\eea
we use the Fourier transform to determine the Green's function in momentum space:
\bea
\bar{G}(\omega)=b\bigg(2i\cos(\pi\Delta)\Gamma(1-2\Delta)\frac{1}{\omega^{1-2\Delta}}\bigg).
\eea
We obtain that
\bea
\bar{\Sigma}(\omega)=J^2b^{3}
\bigg(2i\cos(3\pi\Delta)\Gamma(1-6\Delta)\frac{1}{\omega^{1-6\Delta}}\bigg).
\eea
We then substitute the result into the remaining equation of motion
\bea
\bar{G}\bar{\Sigma}\approx -1.
\eea
The coefficient satisfies
\bea
J^2b^4\pi=\bigg(\frac{1}{2}-\Delta\bigg)\tan(\pi\Delta).
\eea
We get the precise coefficient by using the useful properties of the Gamma function:
\bea
\Gamma(z)\Gamma(-1-z)=-z\Gamma(z)\Gamma(-z)=\frac{\pi }{\sin(\pi z)}.
\eea
Because the SD equations have the reparametrization symmetry, we can select the reparametrization function
\bea
f(\tau)=\tan\bigg(\frac{\pi\tau}{\beta}\bigg)
\label{tan}
\eea
to obtain the thermal Green's function
\bea
\tilde{G}_T(\tau)=b\bigg(\frac{\pi}{\beta\sin\big(\frac{\pi\tau}{\beta}\big)}\bigg)^{2\Delta}\sgn(\tau).
\eea
We have the reparametrization symmetry only in the low-temperature and large-$N$ limits.
Therefore, the result of the thermal Green's function $\tilde{G}_T(\tau)$ only works well for the low-temperature regime.

\subsection{Schwarzian Theory}
\noindent
We can make the shift in the effective action of the SYK model
\bea
\tilde{\Sigma}\rightarrow\tilde{\Sigma}+\partial_{\tau}
\eea
to show only the term breaking the reparametrization symmetry in the effective action \cite{Polchinski:2016xgd,Maldacena:2016hyu}
\bea
\frac{S_{\mathrm{brs}}}{N}=\frac{1}{2}\int d\tau d\tilde{\tau}\ \partial_{\tau}\tilde{G}(\tau, \tilde{\tau})\delta(\tau, \tilde{\tau}).
\eea
We subtract the background term given by the conformal ansatz
\bea
\tilde{G}_c(\tau_1, \tau_2)=\frac{b}{|\tau_1-\tau_2|^{\frac{1}{2}}}\sgn(\tau_1-\tau_2)
\eea
and then the fluctuation shows the Schwarzian theory from the small parameter $|\tau_1-\tau_2|\ll 1$,
\bea
\frac{S_{\mathrm{eff}}}{N}&=&-\frac{1}{2}\int d\tau_1d\tau_2\ \big(\tilde{G}(\tau_1, \tau_2)-\tilde{G}_c(\tau_1, \tau_2)\big)\partial_{\tau_1}\delta(\tau_1-\tau_2)+\cdots
\nn\\
&=&-\frac{b}{48}\int d\tau_2\ Sch(f, \tau_2)\int d\tau_1\ (\tau_1-\tau_2)^{\frac{3}{2}}\sgn(\tau_1-\tau_2)\partial_{\tau_1}\delta(\tau_1-\tau_2)+\cdots
\nn\\
&=&-\frac{b}{48}\int d\tau_2\ Sch(f, \tau_2)\int d\tau\ \tau^{\frac{3}{2}}\sgn(\tau)\partial_{\tau}\delta(\tau)+\cdots,
\eea
where
\bea
\tilde{G}(\tau_1, \tau_2)-\tilde{G}_c(\tau_1, \tau_2)
&=&b\Bigg\lbrack\Bigg(\frac{f^{\prime}(\tau_1)-f^{\prime}(\tau_2)}{\big(f(\tau_1)-f(\tau_2)\big)^2}\Bigg)^{\frac{1}{4}}\sgn\big(f(\tau_1)-f(\tau_2)\big)-\frac{\sgn(\tau_1-\tau_2)}{|\tau_1-\tau_2|^{\frac{1}{2}}}\Bigg\rbrack.
\nn\\
&=&\frac{b}{24}(\tau_1-\tau_2)^{\frac{3}{2}}Sch(f, \tau_2)\sgn(\tau_1-\tau_2)+\cdots.
\eea
One choice of $f(\tau)$ is Eq. \eqref{tan}.
Hence, we obtain the consistent boundary description of the JT gravity from the low-energy limit of the SYK model.
The CFT$_1$ only has the SL(2, $\mathbb{R}$) symmetry.
The temporal reparametrization symmetry is spontaneously broken to the SL(2, $\mathbb{R}$) symmetry in the SYK model.
This phenomenon of spontaneous symmetry breaking engenders a soft mode $f(\tau)$ that facilitates the generation of the Schwarzian action.
Hence, the SYK model provides a more general holographic principle, a nearly AdS$_2$/CFT$_1$ correspondence, because it only approximately realizes the CFT$_1$.
\\

\noindent
To summarize, we introduced the Sachdev-Ye-Kitaev model. 
We explained the SD equation and the low-energy effective theory along the lines of \cite{Polchinski:2016xgd,Maldacena:2016hyu}.
The discussion above can be repeated for other $q\neq 2$ with $\Delta=1/q$, while for $q=2$, the Green's function is exactly conformally invariant.
Therefore, the $q=2$ case does not have spontaneous conformal symmetry breaking to generate the Schwarzian theory.

\section{Quantum Chaos}
\label{sec:4}
\noindent
We first introduce classical chaos in the interval case to explicitly define the conditions under which chaos holds.
We then transition to quantum chaos from the classical chaos analogy \cite{Berry:1977zz}.
The direct connection between the holographic principle and quantum chaos is through the Lyapunov exponent \cite{Larkin:1969,Maldacena:2015waa,Narovlansky:2025tpb}.
We then introduce RMT-based diagnostics of the spectrum, providing a more robust approach to diagnosing quantum chaos in the semiclassical regime \cite{Dyson:1962es,Bohigas:1983er,Muller:2004nb}.

\subsection{Chaos for Interval}
\noindent
The simplest setting is an iterated map on an interval:
\bea
x_{n+1}=f(x_n), \ x_n\in V=[0, 1].
\eea
We choose an initial point $x_0$, then repeatedly apply $f$:
\bea
x_1=f(x_0), \ x_2=f\big(f(x_0)\big), \ \cdots, \ x_n=f^n(x_0).
\eea
Although the phase space is one-dimensional, these systems can already display rich chaotic behavior.
The typical example is the logistic map
\bea
f(x)=4x(1-x).
\eea
The remarkable fact is that one-dimensional maps can be fully deterministic but still unpredictable in practice.
Because knowing $x_0$ only to finite precision means that we really know a small interval of possible initial states.
Under chaos:
\begin{itemize}
\item{this interval is stretched,}
\item{spread across the phase space.}
\end{itemize}
After many iterations, finite-precision uncertainty grows to macroscopic uncertainty.
Hence, the practical unpredictability comes from
\begin{itemize}
\item{determinstic evolution,}
\item{exponential amplification of uncertainty.}
\end{itemize}

\noindent
We say that $f$ is chaotic on $V$ if it satisfies the following three properties:
\begin{itemize}
\item{$f$ has sensitive dependence on initial conditions;
\\
A map $f$ has sensitive dependence on a set $X\in V$ if there exists $\delta>0$ such that for any $x\in X$ and any neighborhood of $x$, one can find $y$ nearby and $n\ge 0$ with
\bea
|f^n(x)-f^n(y)|>\delta.
\eea
This captures the idea that arbitrarily close initial conditions can eventually separate by a finite amount.
}
\item{$f$ is topologically transitive on $V$;
\\
A map $f: V\rightarrow V$ is topologically transitive if for any two non-empty open sets $I, J\in V$, there exists $n\ge 0$ such that
\bea
f^n(I)\cap U\neq \emptyset.
\eea
This means that an orbit starting somewhere in $I$ can eventually reach $J$.
}
\item{periodic points are dense in $V$.
\\
Dense periodic points mean that every open interval contains a periodic point.
}
\end{itemize}
The first and second properties correspond to irregular, unpredictable behavior and hence agree with intuitive ideas about chaos.
The third property, however, seems at odds with the previous two, since we usually think of periodic behavior as regular and predictable.
On the other hand, these periodic points are all unstable, and this means that as soon as an orbit comes close to a periodic point, it will be pushed away somewhere else.
Hence, in practical terms, a dense set of periodic orbits does not imply orderly behavior.
Although sensitivity is generally considered one of the most important ingredients in chaotic behavior, other ingredients are also highly relevant.
Sensitivity refers to the behavior of two orbits that diverge away from each other as iteration proceeds.
On the other hand, the presence of chaos is often inferred from the behavior of a single orbit that wanders throughout the codomain of the mapping (rather than settling into regular, predictable behavior).
The orbit appears "chaotic" when its points are projected onto the vertical line at the right and densely fill the interval.
Hence, we can replace the intuitive idea of a "chaotic orbit" with the precise mathematical concept of a dense orbit.

\subsubsection{Sensitive Dependence is Redundant}
\noindent
Any definition of chaos must face the obvious question: If $f$ is chaotic, and if we have a commutative diagram as in Fig. \ref{commutative diagram}, where $X, Y$ are metric spaces, and $h$ is a homeomorphism, then is $g$ chaotic?
\begin{figure}[t]
\centering
{\huge
\begin{tikzcd}
X \ar[r, "f"] \arrow[d, "h"'] & X \ar[d,"h"] \\
Y \ar[r, "g"] & Y
\end{tikzcd}
}
\caption{The commutative diagram with metric spaces $X$ and $Y$, homeomorphism $h$, and continuous maps $f$ and $g$.}
\label{commutative diagram}
\end{figure}
If we have the following conditions for all $x$, $y$, and $z$ on any set $S$:
\begin{itemize}
\item{$d(x, y)\ge 0$ and $d(x, y)=0$ if and only if $x=y$;}
\item{$d(x, y)=d(y, x)$;}
\item{$d(x, z)\le d(x, y)+d(y, z)$,}
\end{itemize}
the $d$ is called a metric on $S$, and the pair ($S$, $d$) is called a metric space.
Topological transitivity and the existence of dense periodic points are preserved, since they are topological conditions.
However, sensitivity is a metric property and, in general, is not preserved, as the following simple example shows \cite{Banks:1992}.
\\

\noindent
Let $X$ be the subset $(1, \infty)$ of the real line, equipped with the standard metric; let $f$ be multiplication by 2; let $Y$ be the set of positive reals; and let $h$ be $\ln$.
Clearly, $f$ has sensitive dependence on initial conditions, but $g$ is a translation and hence not sensitive with respect to the standard metric on the positive reals.
Let us remark that sensitivity can be regarded as a topological concept if one restricts to compact spaces $X$.
Indeed, suppose that $X$ is compact, that we have the same commutative diagrams, and that $f$ has sensitive dependence on initial conditions; then $g$ also has sensitive dependence on initial conditions.
\\

\noindent
If $X$ is a metric space without isolated points, and $f: X\rightarrow X$ is continuous, transitive, and has periodic points, then $f$ is topologically sensitive.
A point $x\in X$ is called an isolated point if there exists some open neighborhood $U$ of $x$ such that $U\cap X=\{x\}$.
An interval, or any non-trivial interval subset, has no isolated points, so this applies.
We can briefly write this mathematical result as:
\bea
\mathrm{transitivity}+\mathrm{dense\ periodic\ points} \Longrightarrow \mathrm{sensitivity}.
\eea
Hence, the sensitive dependence is redundant.
\\

\noindent
Let us now present the structure of the proof in concrete terms.
We let:
\begin{itemize}
\item{$u$ be a periodic point;}
\item{$q$ be another periodic point (chosen elsewhere in the interval).}
\end{itemize}
Their orbits are:
\begin{itemize}
\item{${\cal O}(u)=\{u, f(u), \cdots, f^{k-1}(u)\}$;}
\item{${\cal O}(q)=\{q, f(q), \cdots, f^{m-1}(q)\}$.}
\end{itemize}
These are finite sets.
Since the orbits are distinct, these sets are disjoint.
Therefore, we define
\bea
\eta\equiv\min\{|x-y|: x\in{\cal O}(u), y\in{\cal O}(q)\}>0.
\eea
This is the key: a strictly positive separation exists.
Now we choose
\bea
\delta=\frac{\eta}{4}.
\eea
Since $u$ is periodic
\bea
f^n(u)\in{\cal O}(u)\ \forall n,
\eea
the orbit of $u$ never leaves this finite set.
Now use topological transitivity: Given any open set $U$, there exists $z\in U$ such that $ z$'s orbit eventually enters any other open set.
Therefore, we can choose $z$ such that for some $n$, $f^n(z)\in$ a small neighborhood of ${\cal O}(q)$.
More concretely: there exists $n$ such that
\bea
|f^n(z)-{\cal O}(q)|<\delta.
\eea
We choose $u$ and $z$ to lie in the arbitrarily small neighborhood $U$, an arbitrarily small neighborhood of $x$.
At time $n$:
\begin{itemize}
\item{$f^n(u)\in{\cal O}(u)$;}
\item{$f^n(z)$ is within $\delta$ of ${\cal O}(q)$.}
\end{itemize}
Since ${\cal O}(u)$ and ${\cal O}(q)$ are at least $\eta=4\delta$ apart, we get:
\bea
|f^n(z)-f^n(u)|\ge \eta-\delta=4\delta-\delta=3\delta.
\eea
Hence, we obtain
\bea
|f^n(z)-f^n(u)|>2\delta.
\eea
By the triangle inequality
\bea
|f^n(z)-f^n(u)|\le|f^n(z)-f^n(x)|+|f^n(u)-f^n(x)|,
\eea
at least one of
\bea
|f^n(z)-f^n(x)|>\delta; \ |f^n(u)-f^n(x)|>\delta
\eea
must hold.
This proves sensitivity.
Hence, for the interval chaos, the real content is topological transitivity plus dense periodic points.
The dense periodic points give rise to many distinct recurrent orbit patterns everywhere, and the topological transitivity forces arbitrarily small neighborhoods to access different parts of the interval; therefore, nearby initial points cannot remain close forever, so a uniform sensitivity scale emerges.

\subsubsection{Sufficient Condition for Interval Chaos}
\noindent
Usually, it is hard to know the sufficient condition for chaos.
However, in the interval case, such a simple situation, we can discover a sufficient condition for chaos.
The $f$ is a one-hump mapping on $\lbrack a, c\rbrack$ if (i) it is continuous, and (ii) it strictly increases from $f(a)=0$ to $f(b)=1$ and then strictly decreases to $f(c)=0$.
Let $f$ defined on $[0,1]$
be a one-hump mapping symmetric about 1/2.
The mapping $f$ satisfies $f(1/2-z)=f(1/2+z)$ for all $z\in$ [0, 1/2].
If $f^{\prime}(0)>1$, and if $f$ has negative Schwarzian derivative, $Sch(f, x)<0$, (except at $x=1/2$), then $f$ has chaotic behavior \cite{Guckenheimer:1979,Milnor:1985}.
Note that
$Sch(f,x)\equiv f'''/f' - (3/2)(f''/f')^2$ is non-positive for any second-order polynomial $f(x)$ of $x$, while $f(x)=x(1-x)$ is not chaotic with $x=0$ being the stable fixed point.
The theorem cannot be applied because $f^{\prime}(0)=1$.
We can apply this theorem to $f$: [0, 1]$\rightarrow$ [0, 1] where $f(x)=4x(1-x)$.
Hence $f^{\prime}(0)=4>1$ while the Schwarzian derivative of $f$ is negative except for $x=1/2$.
Thus, $f$ is chaotic.
This sufficient condition makes it particularly simple to determine whether the mapping is chaotic.

\subsection{Quantum Signatures of Chaos}
\noindent
Quantum chaos studies how signatures of classical chaos manifest in quantum systems, where
\begin{itemize}
\item{Dynamics is given by a linear evolution of a state (=vector in Hilbert space);}
\item{No trajectories exist (no classical phase space paths);}
\item{No exponential divergence of states in Hilbert space.}
\end{itemize}
The core tension is:
\begin{itemize}
\item{Classical chaos$\rightarrow$ trajectories diverge exponentially;}
\item{Quantum mechanics$\rightarrow$ linear evolution, no trajectory divergence.}
\end{itemize}
The classical definition of chaos fails directly in quantum systems.
Therefore, quantum chaos asks: What replaces classical chaos in quantum systems?
Quantum chaos is not about trajectories, but about how quantum states distribute over classical phase structures.
Quantum eigenstates inherit the structure of classical phase space, even though quantum mechanics has no trajectories.
In the semi-classical limit $\hbar\rightarrow 0$, quantum states "resolve" classical phase structures.
We can classify the spectrum (regular and irregular) of quantum eigenstates based on the structure of the corresponding classical phase space and semiclassical mechanics.
The following properties give the irregular spectrum:
\begin{itemize}
\item{There is no unambiguous assignment of a quantum number to a state.}
\item{The discrete bound-state quantum spectrum tends to a continuous classical spectrum in the classical limit.
The energy differences form a discrete distribution that tends to the continuous distribution.
The distribution of levels in the irregular spectrum appears random.}
\item{There are no neighboring states for weak perturbations.
A "neighboring state" with energy $E_0$ is a state with energy $E$ close to $E_0$.
In other words, under weak external perturbations, a state is much more strongly coupled to the neighboring states than to other states in the regular spectrum.}
\end{itemize}
When the spectrum does not satisfy the above properties, the quantum system has regular semi-classical mechanics and a regular spectrum.
To shift from the classification to understand the universal structure of chaotic eigenstates, the next question in quantum chaos is: If the classical system is chaotic, what do the quantum eigenstates look like \cite{Berry:1977zz,Berry:1977wpp}?
Chaotic eigenstates are universal random objects \cite{Berry:1977zz,Berry:1977wpp}.
Chaos in quantum mechanics appears as randomness given by random wavefunctions independent of system details \cite{Berry:1977zz,Berry:1977wpp}.

\subsubsection{Lyapunov Exponent}
\noindent
The OTOC diagnostic is built from two operators, $W$ and $V$, chosen to be simple local operators.
A standard object is
\bea
C(t)\equiv-\langle(\lbrack W(t), V(0)\rbrack)^2\rangle_{\beta},
\eea
where $W(t)=\exp(i Ht)W(0)\exp(-iHt)$, $H$ is the Hamiltonian, and the thermal expectation value is taken at inverse temperature $\beta$.
Equivalently, one often studies the related four-point function
\bea
F(t)=\langle W^{\dagger}(t)V(0)W(t)V(0)\rangle_{\beta}
\eea
or a regulated version with Euclidean insertions space around the thermal circle.
The point is that if $W(t)$ and $V(0)$ nearly commute, then $C(t)$ is small; if time evolution causes $W(t)$ to spread into more degrees of freedom, the commutator grows.
In that sense, OTOCs measure operator growth and scrambling, not just ordinary relaxation.
Historically, a squared commutator-type quantity was used in a superconductivity context \cite{Larkin:1969}.
Modern chaos language came much later.
\\

\noindent
A useful semi-classical intuition is
\bea
\lbrack x(t), p(0)\rbrack\sim i\hbar\frac{\partial x(t)}{\partial x(0)}.
\eea
If the classical sensitivity grows like $\exp(\lambda t)$, then the squared commutator can inherit an early-time exponential regime.
That is the bridge between the classical Lyapunov exponent and the quantum chaos literature.
\\

\noindent
One often writes an early-time scrambling form
\bea
C(t)\sim \frac{1}{N_{\mathrm{eff}}}e^{\lambda_Lt}
\eea
or, equivalently, a deviation of the OTO four-point function from its factorized value,
\bea
F(t)\approx F_{\mathrm{disc}}-\epsilon e^{\lambda_Lt},
\eea
where $\epsilon> 0$, over an intermediate time window.
The coefficient $\lambda_L$ is called the quantum Lyapunov exponent.
It is not guaranteed to exist in every model, and even when it does, the exponential regime is typically only parametric in special limits such as large $N$ or semi-classical limits.
The central general result is the Maldacena-Shenker-Stanford chaos bound
\bea
\lambda_L\le\frac{2\pi}{\beta}
\eea
in units with $\hbar=k_B=1$, under assumptions including thermal equilibrium, suitable mathematical conditions, and simple operators \cite{Maldacena:2015waa,Narovlansky:2025tpb}.
The bound says that no thermal quantum system for simple operators can exhibit parametrically faster OTOC growth than $2\pi/\beta$ \cite{Maldacena:2015waa,Narovlansky:2025tpb}.
Systems that reach the bound are called maximally chaotic \cite{Maldacena:2015waa}.
\\

\noindent
Two important cautions matter here.
First, $\lambda_L$ is not the only timescale.
One also distinguishes the dissipation time $t_d$, associated with ordinary time-ordered decay, from the scrambling time $t_*\sim\lambda_L^{-1}\ln N_{\mathrm{eff}}$, where the OTOC becomes order one.
Second, fast OTOC growth is not synonymous with all notions of thermalization or transport; it is most directly a probe of how perturbations spread through operator space.
\\

\noindent
The OTOC in SYK is computed from the connected four-point function \cite{Polchinski:2016xgd,Maldacena:2016hyu}.
Specifically, the SYK model considers \cite{Polchinski:2016xgd,Maldacena:2016hyu}
\bea
{\cal F}(\tau_1, \tau_2, \tau_3, \tau_4)=\frac{1}{N^2}\sum_{j, k=1}^N\langle\psi_j(\tau_1)\psi_j(\tau_2)\psi_k(\tau_3)\psi_k(\tau_4)\rangle_{\mathrm{conn}}.
\eea
At large $N$, the relevant diagrams are ladder diagrams \cite{Polchinski:2016xgd,Maldacena:2016hyu}.
\\

\noindent
Now couple a bilocal source $B(x, y)$ to the fermion bilinear
\bea
S_{B}=S_{\mathrm{SYK}}-\int dxdy\ B(x, y)\psi(x)\psi(y),
\eea
where $S_{\mathrm{SYK}}$ is the action of the SYK model.
The bilocal source is anti-symmetric
\bea
B(x, y)=-B(y, x).
\eea
Therefore, we do not double-count the pair $(x, y)$ and $(y, x)$.
Then the generating function is
\bea
Z[B]=\int{\cal D}\psi\ e^{-S_B[\psi]}=\int{\cal D}\psi\ \exp\bigg(S_{\mathrm{SYK}}-\frac{1}{N}\sum_{k=1}^N\int dxdy\ B(x, y)\psi_k(x)\psi_k(y)\bigg).
\eea
Define expectation values in the presence of $B$ by
\bea
\langle{\cal O}\rangle_B=\frac{1}{Z[B]}\int{\cal D}\psi\ {\cal O}e^{-S_B[\psi]}.
\eea
The source-dependent two-point function is
\bea
G_B(x_1, x_2)=\frac{1}{N}\sum_{j=1}^N\langle\psi_j(x_1)\psi_j(x_2)\rangle_B.
\eea
Differentiate $Z[B]$ with respect to $B(x_3, x_4)$.
Since $B$ appears linearly in the exponent, we obtain
\bea
\frac{\delta Z[B]}{\delta B(x_3, x_4)}=\frac{1}{N}\sum_{j=1}^N\int{\cal D}\psi\ \psi_j(x_3)\psi_j(x_4)e^{-S_B[\psi]}.
\eea
Therefore, we get
\bea
\frac{\delta\ln Z[B]}{\delta B(x_3, x_4)}=\frac{1}{Z[B]}\frac{\delta Z[B]}{\delta B(x_3, x_4)}=\frac{1}{N}\sum_{j=1}^N\langle\psi_j(x_3)\psi_j(x_4)\rangle_B.
\eea
Hence, we get
\bea
\frac{\delta\ln Z[B]}{\delta B(x_3, x_4)}=G_B(x_3, x_4).
\eea
\\

\noindent
Now let
\bea
G_B(x_1, x_2)=\frac{1}{Z[B]}\frac{1}{N}\sum_{j=1}^N\int{\cal D}\psi\ \psi_j(x_1)\psi_j(x_2)e^{-S_B[\psi]}.
\eea
Differentiate with respect to $B(x_3, x_4)$,
\bea
\frac{\delta G_B(x_1, x_2)}{\delta B(x_3, x_4)}=\frac{\delta}{\delta B(x_3, x_4)}\bigg(\frac{1}{Z[B]}\frac{1}{N}\sum_{j=1}^N\int{\cal D}\psi\ \psi_j(x_1)\psi_j(x_2)e^{-S_B}\bigg).
\eea
Use the product rule.
There are two contributions:
\begin{itemize}
\item{derivative of the numerator;}
\item{derivative of $1/Z[B]$.}
\end{itemize}
We get
\bea
&&
\frac{1}{N}\sum_{j=1}^N\frac{\delta}{\delta B(x_3, x_4)}\int{\cal D}\psi\ \psi_j(x_1)\psi_j(x_2)e^{-S_B}
\nn\\
&=&\frac{1}{N^2}\sum_{j, k=1}^N\int{\cal D}\psi\ \psi_j(x_1)\psi_j(x_2)\psi_k(x_3)\psi_k(x_4)e^{-S_B}.
\eea
Since we have
\bea
\frac{\delta}{\delta B}\bigg(\frac{1}{Z}\bigg)=-\frac{1}{Z^2}\frac{\delta Z}{\delta B},
\eea
this contributes
\bea
-\frac{1}{(Z[B])^2}\bigg(\frac{\delta Z[B]}{\delta B(x_3, x_4)}\bigg)\frac{1}{N}\sum_{j=1}^N\bigg(\int{\cal D}\psi\ \psi_j(x_1)\psi_j(x_2)e^{-S_B}\bigg).
\eea
Adding the two pieces,
\bea
&&
\frac{\delta G_B(x_1, x_2)}{\delta B(x_3, x_4)}
\nn\\
&=&\sum_{j, k=1}^N\big(\langle\psi_j(x_1)\psi_j(x_2)\psi_k(x_3)\psi_k(x_4)\rangle_B
-
\langle\psi_j(x_1)\psi_j(x_2)\rangle_B\langle\psi_k(x_3)\psi_k(x_4)\rangle_B\big).
\eea
At $B=0$, this becomes
\bea
\frac{\delta G_B(x_1, x_2)}{\delta B(x_3, x_4)}\bigg|_{B=0}=\frac{1}{N^2}\sum_{j ,k=1}^N\langle\psi_j(x_1)\psi_j(x_2)\psi_k(x_3)\psi_k(x_4)\rangle_{\mathrm{conn}}
={\cal F}(x_1, x_2, x_3, x_4).
\eea
\\

\noindent
In the SYK model at the IR limit, the two-point function in the presence of the source satisfies
\bea
G_B^{-1}=-2B-\Sigma[G_B].
\eea
We now differentiate the equation with respect to $B(x_3, x_4)$,
\bea
\frac{\delta G_B^{-1}}{\delta B(x_3, x_4)}=-2\frac{\delta B}{\delta B(x_3, x_4)}-\frac{\delta\Sigma[G_B]}{\delta B(x_3, x_4)}.
\eea
Now use the chain rule on the self-energy
\bea
\frac{\delta\Sigma(x_a, x_b)}{\delta B(x_3, x_4)}=\int dx_5dx_6\ \frac{\delta\Sigma(x_a, x_b)}{\delta G_B(x_5, x_6)}\frac{\delta G_J(x_5, x_6)}{\delta B(x_3, x_4)},
\eea
but by definition
\bea
\frac{\delta G_B(x_5, x_6)}{\delta B(x_3, x_4)}={\cal F}(x_5, x_6, x_3, x_4).
\eea
Hence, we obtain
\bea
\frac{\delta\Sigma(x_a, x_b)}{\delta B(x_3, x_4)}=\int dx_5dx_6\ \frac{\delta\Sigma(x_a, x_b)}{\delta G_B(x_5, x_6)}{\cal F}(x_5, x_6, x_3, x_4).
\eea
\\

\noindent
Now use the basic identity
\bea
\delta G_B=-G_B(\delta G_B^{-1})G_B.
\eea
Therefore, we obtain
\bea
\frac{\delta G_B}{\delta B(x_3, x_4)}=-G_B\frac{\delta G_B^{-1}}{\delta B(x_3, x_4)}G_B
=2G_B\frac{\delta B}{\delta B(x_3, x_4)}G_B+G_B\frac{\delta\Sigma[G_B]}{\delta B(x_3, x_4)}G_B.
\eea
Now rewrite this in coordinates
\bea
&&
{\cal F}(x_1, x_2, x_3, x_4)
\nn\\
&=&2\int dx_adx_b\ G_B(x_1, x_a)\frac{\delta B(x_a, x_b)}{\delta B(x_3, x_4)}G_B(x_b, x_2)
\nn\\
&&
+\int dx_adx_b\ G_B(x_1, x_a)\frac{\delta\Sigma(x_a, x_b)}{\delta B(x_3, x_4)}G_B(x_b, x_2).
\eea
Let us define the first term as ${\cal F}_0$,
\bea
{\cal F}_0(x_1, x_2, x_3, x_4)\equiv2\int dx_adx_b\ G_B(x_1, x_a)\frac{\delta B(x_a, x_b)}{\delta B(x_3, x_4)}G_B(x_b, x_2).
\eea
Now consider the second term and insert the chain rule expression:
\bea
&&
\int dx_adx_b\ G_B(x_1, x_a)\frac{\delta\Sigma(x_a, x_b)}{\delta B(x_3, x_4)}G_B(x_b, x_2)
\nn\\
&=&\int dx_adx_bdx_5dx_6\ G_B(x_1, x_a)\frac{\delta\Sigma(x_a, x_b)}{\delta G_B(x_5, x_6)}{\cal F}(x_5, x_5, x_3, x_4)G_B(x_b, x_2)
\nn\\
&\equiv&\int dx_5dx_6\ K(x_1, x_2, x_5, x_6){\cal F}(x_5, x_6, x_3, x_4),
\eea
where the kernel is defined as
\bea
K(x_1, x_2, x_5, x_6)\equiv\int dx_adx_b\ G_B(x_1, x_a)\frac{\delta\Sigma(x_a, x_b)}{\delta G_B(x_5, x_6)}G_B(x_b, x_2).
\eea
When the source is zero, we obtain the full equation \cite{Polchinski:2016xgd,Maldacena:2016hyu}
\bea
{\cal F}(x_1, x_2, x_3, x_4)={\cal F}_0(x_1, x_2, x_3, x_4)+\int dx_5dx_6\ K(x_1, x_2, x_5, x_6){\cal F}(x_5, x_6, x_3, x_4),
\nn\\
\eea
and the kernel also satisfies
\bea
\delta G(x_1, x_2)=\int dx_3dx_4\ K(x_1, x_2, x_3, x_4)\delta G(x_3, x_4),
\eea
where the kernel takes the standard bilocal form \cite{Polchinski:2016xgd,Maldacena:2016hyu}
\bea
K(x_1, x_2, x_3, x_4)=-3J^2G(x_{13})G(x_{24})\big(G(x_{34})\big)^2,
\eea
where $x_{ab}\equiv x_a-x_b$.
The connected four-point function can be written schematically as a geometric series \cite{Polchinski:2016xgd,Maldacena:2016hyu}
\bea
{\cal F}={\cal F}_0+K{\cal F}_0+K^2{\cal F}_0+\cdots=(1-K)^{-1}{\cal F}_0.
\eea
Therefore, once we know the eigenfunctions and eigenvalues of $K$, we know the four-point function \cite{Polchinski:2016xgd,Maldacena:2016hyu}.
This is usually solved using conformal eigenfunctions.
\\

\noindent
In the conformal limit, the kernel is diagonalized by conformal three-point-function-like eigenfunctions \cite{Polchinski:2016xgd,Maldacena:2016hyu}
\bea
\int d\tau_3d\tau_4\ K(\tau_1, \tau_2, \tau_3, \tau_4)\Psi_h(\tau_3, \tau_4, \tau_0)=k(h)\Psi_h(\tau_1, \tau_2, \tau_0),
\eea
where
\bea
\Psi_h\propto \frac{\sgn(\tau_{12})}{|\tau_{12}|^{2\Delta-h}|\tau_{10}|^h|\tau_{20}|^h}\sim \langle{\cal O}_{\Delta}(\tau_1){\cal O}_{\Delta}(\tau_2){\cal O}_h(\tau_0)\rangle.
\eea
The ${\cal O}_{\Delta}$ is the CFT operator with the conformal dimension $\Delta$.
One finds eigenvalues $k(h)$ depending on the exchanged conformal weight $h$.
Since the eigenvalues of the kernel do not depend on $\tau_0$, we can take it to infinity to obtain the eigenvalues
\bea
k(h)=\int_{-\infty}^{\infty}d\tau_1d\tau_2\ K(1, 0, \tau_1, \tau_2)\frac{\sgn(\tau_{12})}{|\tau_{12}|^{2\Delta-h}}.
\eea
To build the four-point function, one glues two such three-point structures together by integrating over $\tau_0$.
Specifically, we use
\bea
\Psi_h(\tau_1, \tau_2, \tau_3, \tau_4)\sim\int d\tau_0\ \langle{\cal O}_{\Delta}(\tau_1){\cal O}_{\Delta}(\tau_2){\cal O}_h(\tau_0)\rangle\langle{\cal O}_{1-h}(\tau_0){\cal O}_{\Delta}(\tau_3){\cal O}_{\Delta}(\tau_4)\rangle.
\eea
Because the result is conformally invariant, it can only depend on the cross ratio
\bea
\chi=\frac{\tau_{12}\tau_{34}}{\tau_{13}\tau_{24}}.
\eea
Hence, the eigenfunction becomes a one-variable function $\Psi_h(\chi)$.
\\

\noindent
Because the kernel $K$ is diagonal on $\Psi_h$,
\bea
K\Psi_h(\chi)=k(h)\Psi_h(\chi),
\eea
we expand
\bea
{\cal F}(\chi)=\int dh\ c(h)\Psi_h(\chi)
\eea
plus any discrete contributions if needed.
Similarly, we have:
\bea
{\cal F}_0(\chi)=\int dh\ c_0(h)\Psi_h; \ K{\cal F}(\chi)=\int dh\ c(h)J\Psi_h=\int dh\ c(h)k(h)\Psi_h.
\eea
We then get
\bea
\int dh\ c(h)\Psi_h=\int dh\ c_0(h)\Psi_h+\int dh\ c(h)k(h)\Psi_h,
\eea
and then group terms
\bea
\int dh\ \big(c(h)-c(h)k(h)-c_0(h)\big)\Psi_h=0
\eea
Since $\Psi_h$ are independent, we obtain
\bea
c(h)\big(1-k(h)\big)=c_0(h).
\eea
We then get
\bea
c(h)=\frac{c_0(h)}{1-k(h)}.
\eea
Hence, the conformal four-point function has the spectral form
\bea
{\cal F}(\chi)=\int dh\ \frac{c_0(h)}{1-k(h)}\Psi_h(\chi).
\eea
\\

\noindent
In conformal SYK, we have poles in
\bea
k(h)=1.
\eea
This means that the kernel has an eigenfunction with eigenvalue 1 in the $h$ channel, so the naive ladder sum $1/\big(1-k(h)\big)$ diverges, providing the dominant contributions to ${\cal F}(\chi)$.
The equation $k(h)=1$ has an infinite tower of solutions.
The $k(h)=1$ means the naive conformal four-point function becomes singular in the $h$ channel \cite{Polchinski:2016xgd,Maldacena:2016hyu}.
To get a finite answer, we must include the leading symmetry-breaking correction coming from the UV kinetic term $\partial_{\tau}$ that was neglected in the strict IR saddle \cite{Polchinski:2016xgd,Maldacena:2016hyu}.
This lifts the zero mode slightly \cite{Polchinski:2016xgd,Maldacena:2016hyu}.
This is why the correct low-energy theory is not "pure conformal SYK", but "conformal SYK plus a softly lifted reparametrization mode" \cite{Polchinski:2016xgd,Maldacena:2016hyu}.
There is a mode at $h=2$ for which the conformal kernel eigenvalue reaches
\bea
k(h=2)=1
\eea
in the strict infrared conformal limit \cite{Polchinski:2016xgd,Maldacena:2016hyu}.
The contributions from other $h$ are regular, but the $h=2$ contirbution is enhanced because the IR behavior of ${\cal F}(\chi)$ is dominated by $h=2$ \cite{Polchinski:2016xgd,Maldacena:2016hyu}:
\bea
1-k(h=2)\sim\beta J\gg1; \ 1-k(h\neq 2)\sim O(1).
\eea
The $h=2$ mode is parametrically enhanced by $\beta J$, while the others are not \cite{Polchinski:2016xgd,Maldacena:2016hyu}.
This is not an accident.
The lift of the $h=2$ mode produces the Schwarzian effective action \cite{Polchinski:2016xgd,Maldacena:2016hyu}.
That lifted mode controls the leading OTOC growth.
It is the manifestation of the emergent reparametrization symmetry and its soft breaking \cite{Polchinski:2016xgd,Maldacena:2016hyu}.
Physically, the conformal approximation provides the deformation, the reparametrization mode \cite{Polchinski:2016xgd,Maldacena:2016hyu}.
\\

\noindent
The OTOC is not just any four-point function; it is an analytically continued one, with operators placed on different segments of a complex-time contour.
In practice, one begins from the Euclidean answer and continues
\bea
\tau\rightarrow it+\epsilon
\eea
with small imaginary regulators chosen to enforce the out-of-time ordering \cite{Polchinski:2016xgd,Maldacena:2016hyu}.
In that Lorentzian continuation, the same kernel problem appears. However, the relevant kernel is now the retarded kernel rather than the Euclidean one \cite{Polchinski:2016xgd,Maldacena:2016hyu}.
The exponentially growing piece comes from an eigenfunction of that retarded kernel with eigenvalue 1 \cite{Polchinski:2016xgd,Maldacena:2016hyu}.
Hence, we have three descriptions of the same IR physics:
\begin{itemize}
\item{Euclidean language: the four-point function is nearly singular in the $h=2$ channel.}
\item{Lorentzian language: the retarded kernel has a growing eigenmode.}
\item{Effective-theory language: there is a soft reparametrization mode.}
\end{itemize}
In the conformal/strong-coupling limit, the $h=2$ mode becomes the reparametrization soft mode and dominates the largest exponent of ${\cal F}(t)$,
\bea
{\cal F}(t)\sim\frac{1}{N}e^{\lambda_Lt},
\eea
where
\bea
\lambda_L=\frac{2\pi}{\beta},
\eea
so SYK is maximally chaotic in that regime \cite{Polchinski:2016xgd,Maldacena:2016hyu}.
This is one of the main reasons SYK became the benchmark model for quantum many-body chaos.
More concretely, there is a window
\bea
t_d\ll t\ll t_*,
\eea
in which the OTOC deviation grows exponentially.
The prefactor is suppressed by $1/N$, so the scrambling time scales like
\bea
t_*\sim \frac{\beta}{2\pi}\ln N
\eea
up to model-dependent constants.
This mirrors the "fast scrambler" expectation for black holes.
\\

\noindent
OTOCs are often described as measuring "chaos", but more precisely, they diagnose the growth of non-commutativity under Heisenberg evolution, or equivalently, the spread of initially simple operators into complicated ones.
This makes them especially useful for scrambling, which is stronger than the simple decay of two-point functions.
A system can have short relaxation times without exhibiting a large, clean exponential OTOC window.
Conversely, one should avoid overinterpreting $\lambda_L$.
It is not a universal order parameter for all quantum chaos, and it may be absent, ambiguous, or cutoff-sensitive in finite systems or outside controlled limits.
Therefore, the OTOCs provide an intuitive understanding of quantum chaos, but they do not capture all its features.
In many systems, the OTOC grows but not with a long exponential regime.
The cleanest setting is the large-$N$ limit, where factorization and a parametric window make the exponential piece well-defined.
Another subtlety is regularization.
Different operator orderings and thermal insertions can produce related but not identical correlators, and the chaos bound is proved for appropriately regulated OTOCs with the required analyticity properties \cite{Maldacena:2015waa}.

\subsubsection{Random Matrix Theory}
\noindent
Random Matrix Theory (RMT) studies statistical properties of matrices whose entries are random variables \cite{Dyson:1962es}.
The core idea is:
\begin{itemize}
\item{Replace a complicated system (e.g., nuclei, chaotic Hamiltonians, SYK \cite{Polchinski:2016xgd,Maldacena:2016hyu});}
\item{With an ensemble of matrices with the same symmetry constraints \cite{Dyson:1962es};}
\item{Extract universal statistical behavior \cite{Dyson:1962es}.}
\end{itemize}
RMT predicts that many systems share the same spectral statistics regardless of microscopic details.
This is why:
\begin{itemize}
\item{Nuclear spectra;}
\item{Quantum chaotic systems;}
\item{SYK model}
\end{itemize}
all exhibit similar statistics for each symmetry sector.
Symmetries classify the three fundamental ensembles \cite{Dyson:1962es}:
\begin{itemize}
\item{Gaussian Orthogonal Ensemble (GOE)
\begin{itemize}
\item{Real symmetric matrices}
\item{Time-reversal symmetry}
\end{itemize}
}
\item{Gaussian Unitary Ensemble (GUE)
\begin{itemize}
\item{Complex Hermitian matrices}
\item{No time-reversal symmetry}
\end{itemize}
}
\item{Gaussian Symplectic Ensemble (GSE)
\begin{itemize}
\item{Quaternionic Hermitian matrices}
\item{Time-reversal symmetry}
\end{itemize}
}
\end{itemize}
After diagonalization, the joint probability distributions \cite{Dyson:1962es}
\bea
P(\lambda_j)\propto\prod_{j<k}|\lambda_j-\lambda_k|^{\beta}\prod_{l=1}^ne^{-\frac{\beta}{4}\lambda_k^2},
\eea
where $\lambda_1, \lambda_2, \cdots, \lambda_n$ are the eigenvalues of GUE/GOE/GSE, which are classified by the Dyson index
\bea
\beta=\left\{\begin{array}{ll}
1 & \mbox{GOE}; \\
2 & \mbox{GUE}; \\
4 & \mbox{GSE}.
\end{array} \right.
\eea
This parameter $\beta$ controls the level repulsion.
Bohigas–Giannoni–Schmit studied the Sinai billiard and found that the eigenvalue statistics match RMT predictions and further proposed the BGS conjecture \cite{Bohigas:1983er}.
The BGS conjecture states: Quantum systems whose classical counterparts are chaotic exhibit universal spectral statistics described by Random Matrix Theory (RMT) \cite{Bohigas:1983er}.
In a precise sense, the spectral statistics of a quantum Hamiltonian for each symmetry sector are described by RMT (GOE/GUE/GSE) if the classical limit is chaotic \cite{Bohigas:1983er}.
Because the quantum system lacks trajectories, it lacks a direct definition of chaos.
The BGS conjecture provides a spectral definition of quantum chaos.
We can examine spectral measures such as the level-spacing distribution \cite{Bohigas:1983er}, the averaged adjacent-gap ratio \cite{Bohigas:1983er}, and the spectral form factor \cite{Brezin:1997rze} to assess spectral statistics.
\\

\noindent
The chaotic dynamics give rise to highly complex wavefunctions \cite{Berry:1977zz,Berry:1977wpp}.
The Hamiltonian effectively behaves like a random matrix.
Therefore, the eigenfunctions behave like random superpositions \cite{Berry:1977zz,Berry:1977wpp}.
Through the Gutzwiller trace formula, we can bridge the spectrum to the classical periodic orbits to prove the BGS conjecture in the semi-classical regime \cite{Muller:2004nb}.
\\

\noindent
Given an ordered spectrum:
\bea
E_1<E_2<\cdots,
\eea
and define nearest-neighbor spacings
\bea
\delta_n=E_{n+1}-E_n
\eea
with the rescaling (or unfolding process)
\bea
s_n\equiv\frac{\delta_n}{\langle \delta\rangle},
\eea
so that
\bea
\langle s_n\rangle=1.
\eea
The level spacing distribution is the probability distribution of the normalized nearest-neighbor spacings, $P(s)=\mathrm{Prob}(\mathrm{spacing}=s)$.
This is one of the most fundamental diagnostics of spectral statistics.
If eigenvalues are uncorrelated, the spectral statistics behave as the Poisson distribution (integrable systems)
\bea
P(s)=e^{-s}.
\eea
There is no level repulsion, and the spacings are independent.
The RMT's spectral statistics approximately exhibit the Wigner-Dyson distribution (chaotic systems)
\bea
P(s)\sim s^{\beta}e^{-c s^2},
\eea
where $c$ is a positive constant.
Eigenvalues behave like repelling particles:
\bea
P(s\rightarrow 0)=0.
\eea
The Wigner-Dyson distribution has the exact result for 2$\times$2 matrices, highly accurate for the large matrices \cite{Dyson:1962es}:
\begin{itemize}
\item{GOE
\\
\bea
P(s)=\frac{\pi}{2}se^{-\frac{\pi}{4}s^2}
\eea
}
\item{GUE
\\
\bea
P(s)=\frac{32}{\pi^2}s^2e^{-\frac{4}{\pi}s^2}
\eea
}
\item{GSE
\\
\bea
P(s)=\frac{2^{18}}{3^6\pi^3}s^4e^{-\frac{64}{9\pi}s^2}
\eea
}
\end{itemize}

\noindent
We can define the adjacent gap ratio through the nearest-neighbor spacings
\bea
r_n=\frac{\min(\delta_n, \delta_{n-1})}{\max(\delta_n, \delta_{n-1})}.
\eea
The $r_n$ lies in the range
\bea
0\le r_n\le 1.
\eea
Another observer of the RMT is to average the adjacent gap ratio
\bea
\langle r\rangle=\frac{1}{D}\sum_{n=1}^Dr_n.
\eea
We can use the averaged gap ratio to distinguish the distributions quantitatively \cite{Oganesyan:2007wpd,Atas:2013gvn,Nishigaki:2024yjr}:
\begin{itemize}
\item{Poisson: $\langle r\rangle\approx 0.386$;}
\item{GOE: $\langle r\rangle\approx 0.5307$;}
\item{GUE: $\langle r\rangle\approx 0.5996$;}
\item{GSE: $\langle r\rangle\approx 0.6762$.}
\end{itemize}
In the SYK model, each symmetry sector of the Hamiltonian can be classified by the number of Majorana fermions and the charge-conjugation operators when $N\gg 1$ \cite{You:2016ldz,Cotler:2016fpe}:
\begin{itemize}
\item $N/2 \bmod 4 = 0, ({\cal P}^2, {\cal R}^2) = (1, 1):$ no degeneracy, GOE;
\item $N/2 \bmod 4 = 1, ({\cal P}^2, {\cal R}^2) = (1, -1):$ no degeneracy, GUE;
\item $N/2 \bmod 4 = 2, ({\cal P}^2, {\cal R}^2) = (-1, -1):$ twofold degeneracy, GSE;
\item $N/2 \bmod 4 = 3, ({\cal P}^2, {\cal R}^2) = (-1, 1):$ no degeneracy, GUE.
\end{itemize}
The charge-conjugation operators are:
\bea
{\cal P}\equiv K\prod_{j=1}^{\frac{N}{2}}\gamma_{2j-1}; \ {\cal R}\equiv K\prod_{j=1}^{\frac{N}{2}}i\gamma_{2j},
\eea
where $\gamma_j$ are the Majorana operators, and $K$ is an anti-linear operator and acts as complex conjugation.
\\

\noindent
Given a Hamiltonian $H$ with eigenvalues $E_n$, we can define the partition function
\bea
Z(\beta, t)\equiv\mathrm{Tr}\bigg(e^{-(\beta+it)H}\bigg).
\eea
The spectral form factor is
\bea
K(t)=\big\langle|Z(\beta, t)|^2\big\rangle.
\eea
At infinite temperature ($\beta=0$), we get
\bea
K(t)=\bigg\langle\sum_{m, n}e^{-i(E_m-E_n)t}\bigg\rangle=\sum_n 1+\sum_{m\neq n}e^{-i(E_m-E_n)t}.
\eea
We observe the disconnected term
\bea
\big\vert\langle Z(\beta,t)\rangle\big\vert^2
\eea
and the connected term
\bea
\big\langle|Z(\beta, t)|^2\big\rangle-\big\vert\langle Z(\beta,t)\rangle\big\vert^2.
\eea
The spectral form factor measures:
\begin{itemize}
\item{correlations between eigenvalues;}
\item{interference between energy levels;}
\item{quantum chaos signatures.}
\end{itemize}
The spectral form factor measures the long-range (global) correlations. However, the averaged gap ratio measures the short-range (local) correlations.
The density of states is
\bea
\rho(E)=\sum_n\delta(E-E_n).
\eea
The partition function can be rewritten as
\bea
Z(t)=\int dE\ \rho(E)e^{-iEt}.
\eea
The spectral form factor is the Fourier transform of the two-point spectral correlation function
\bea
K(t)=\bigg\langle\int dEdE^{\prime}\ \rho(E)\rho(E^{\prime})e^{-i(E-E^{\prime})t}\bigg\rangle.
\eea
The hallmark of quantum chaos can be understood from the three regimes of the spectral form factor:
\begin{itemize}
\item{Slope
\begin{itemize}
\item{Early time decay}
\end{itemize}
}
\item{Linear Ramp
\begin{itemize}
\item{Linear growth}
\item{Signature of level repulsion}
\end{itemize}
}
\item{Plateau
\begin{itemize}
\item {Saturation at late times}
\item{Finite Hilbert space dimension}
\end{itemize}
}
\end{itemize}
The above quantum chaos diagnostic, indeed, requires the thermodynamic limit to restrict the study of many-body quantum chaos.
In the few-body case, we do not have a measure that yields a universal result for distinguishing integrable from non-integrable systems.
The SYK model incorporates both integrable and non-integrable configurations via the disorder parameters \cite{Lau:2020qnl,Ozaki:2025mma}.
We cannot use the conserved charge to define integrability for a model with disorder parameters.
However, the spectral statistics seem to provide an effective diagnostic, as in the model without disorder parameters.

\section{Matter Coupling from SYK Model}
\label{sec:5}
\noindent
We first introduce the AdS/CFT dictionary for establishing the relationship between the bulk gravity and the boundary field theory \cite{Maldacena:1997re,Gubser:1998bc,Witten:1998qj}.
We then start applying the relation to construct the expected models consistent with the boundary description of JT gravity with a free scalar field and introduce the interacting terms by introducing a non-Gaussian distribution to the disorder parameters \cite{Lau:2023pot,Lau:2025dgd}.
In the end, we discuss the RMT result from the matter coupling \cite{Lau:2023pot,Lau:2025dgd}.

\subsection{AdS/CFT Dictionary}
\noindent
The AdS/CFT "dictionary" means a set of translation rules \cite{Maldacena:1997re,Gubser:1998bc,Witten:1998qj}:
\begin{itemize}
\item{bulk fields in AdS correspond to boundary operators,}
\item{bulk masses correspond to conformal dimensions,}
\item{bulk gauge symmetries correspond to global symmetries in CFT,}
\item{bulk partition functions with fixed boundary conditions correspond to generating functionals in CFT,}
\item{bulk black holes correspond to thermal states in CFT,}
\item{bulk radial direction corresponds to energy scale in CFT.}
\end{itemize}
The dictionary tells us how to translate statements from the gravity side to the field theory side.
A standard Poincaré patch metric for AdS$_{d+1}$ is
\bea
ds^2=\frac{L^2}{z^2}(dz^2+\eta_{\mu\nu}dx^{\mu}dx^{\nu}),
\eea
where $z>0$.
Here:
\begin{itemize}
\item{$L$ is the AdS radius,}
\item{$x^{\mu}$, $\mu=0, 1, \cdots, d-1$, are boundary coordinates,}
\item{$z$ is radial coordinate,}
\item{the boundary is at $z\rightarrow 0$.}
\end{itemize}
The bulk is $(x^{\mu}, z)$, and the boundary is only $x^{\mu}$.
The CFT lives on the conformal class of the boundary metric \cite{Maldacena:1997re,Gubser:1998bc,Witten:1998qj}.
In Poincaré coordinates, the boundary metric is just Minkowski space up to a Weyl factor.
\\

\noindent
The isometry group of AdS$_{d+1}$ is SO(2, $d$).
The conformal group of CFT$_d$ is also SO(2, $d$).
This is already a strong hint that the two theories are equivalent \cite{Maldacena:1997re,Gubser:1998bc,Witten:1998qj}.
More specifically \cite{Maldacena:1997re,Gubser:1998bc,Witten:1998qj}:
\begin{itemize}
\item{bulk time translations$\leftrightarrow$ CFT Hamiltonian,}
\item{bulk spatial rotations$\leftrightarrow$ boundary rotations,}
\item{radial rescalings in AdS$\leftrightarrow$ dilatations in CFT.}
\end{itemize}

\noindent
The central formula of the dictionary is the Gubser-Klebanov-Polyakov-Witten (GKPW) relation \cite{Maldacena:1997re,Gubser:1998bc,Witten:1998qj}
\bea
Z_{\mathrm{bulk}}[\phi_0(x)]=\bigg\langle\exp\bigg(\int d^dx\ \phi_0(x){\cal O}(x)\bigg)\bigg\rangle_{\mathrm{CFT}}.
\eea
Here:
\begin{itemize}
\item{$\phi(x, z)$ is a bulk field,}
\item{$\phi_0(x)$ is its boundary field,}
\item{${\cal O}(x)$ is the dual CFT operator.}
\end{itemize}
This means \cite{Maldacena:1997re,Gubser:1998bc,Witten:1998qj}:
\begin{itemize}
\item{the boundary value of the bulk field acts as a source for the dual operator in CFT,}
\item{differentiating the bulk partition function with respect to $\phi_0(x)$ gives boundary correlators of ${\cal O}(x)$.}
\end{itemize}
In the classical gravity limit \cite{Maldacena:1997re,Gubser:1998bc,Witten:1998qj}
\bea
Z_{\mathrm{bulk}}[\phi_0]\approx e^{-S_{\mathrm{bulk, \ on-shell}}[\phi_0]}
\eea
in Euclidean signature, so that \cite{Maldacena:1997re,Gubser:1998bc,Witten:1998qj}
\bea
\bigg\langle\exp\bigg(\int d^dx\ \phi_0{\cal O}\bigg)\bigg\rangle\approx e^{-S_{\mathrm{on-shell}}[\phi_0]}.
\eea
This is the most-used practical dictionary entry.
The most basic entry is \cite{Maldacena:1997re,Gubser:1998bc,Witten:1998qj}:
\begin{itemize}
\item{bulk scalar $\phi$$\longleftrightarrow$ scalar operator ${\cal O}$,}
\item{bulk gauge field $A_{M}$$\longleftrightarrow$ conserved current $J_{\mu}$,}
\item{bulk metric fluctuation $h_{MN}$$\longleftrightarrow$ stress tensor $T_{\mu\nu}$,}
\item{bulk fermion $\psi$$\longleftrightarrow$fermionic operator ${\cal O}_{\psi}$.}
\end{itemize}
The spin and symmetry properties of the bulk field match those of the boundary operator \cite{Maldacena:1997re,Gubser:1998bc,Witten:1998qj}.
\\

\noindent
For a bulk scalar field $\phi(z, x)$ of mass $m$, near the boundary $z\rightarrow 0$ \cite{Maldacena:1997re,Gubser:1998bc,Witten:1998qj},
\bea
\phi(z, x)\sim z^{d-\Delta}\phi_{(0)}(x)+z^{\Delta} A(x)+\cdots,
\eea
where $\Delta$ satisfies \cite{Maldacena:1997re,Gubser:1998bc,Witten:1998qj}
\bea
\Delta(\Delta-d)=m^2L^2.
\eea
Equivalently,
\bea
\Delta=\frac{d}{2}\pm\sqrt{\frac{d^2}{4}+m^2L^2}.
\eea
The coefficient $\phi_{(0)}(x)$ is the source of ${\cal O}(x)$, and the coefficient $A(x)$ is related to the expectation value $\langle {\cal O}(x)\rangle$ \cite{Maldacena:1997re,Gubser:1998bc,Witten:1998qj}.
Therefore, one mode is the source, and the other is the response.
Once we know the on-shell action as a functional of the source $\phi_{(0)}$, we compute the correlators by differentiation \cite{Maldacena:1997re,Gubser:1998bc,Witten:1998qj}
\bea
\langle{\cal O}(x){\cal O}(y)\rangle=\frac{\delta^2S_{\mathrm{on-shell}}}{\delta\phi_{(0)}(x)\delta\phi_{(0)}(y)}.
\eea
For a scalar operator of conformal dimension $\Delta$, conformal symmetry gives
\bea
\langle{\cal O}(x){\cal O}(y)\rangle\propto \frac{1}{|x-y|^{2\Delta}}.
\eea
This result is reproduced by the AdS calculation \cite{Maldacena:1997re,Gubser:1998bc,Witten:1998qj}.

\subsection{Scalar Fields}
\noindent
We consider the massless scalar fields, and the action is
\bea
S_{\mathrm{sc}}=\frac{\lambda}{32\pi G_2}\sum_{k=1}^M\int d^2x\sqrt{|\det{g_{\rho\sigma}}|}\ (\nabla_{\mu}\Phi_k)(\nabla^{\mu}\Phi_k),
\eea
where $M$ is the number of scalar fields $\Phi_k$, and $\lambda$ is the matter coupling.
Because the scalar fields satisfy
\bea
\partial_+\partial_-\Phi_k=0,
\eea
The general solution for $\Phi_k$ is
\bea
\Phi_k=\frac{1}{2\pi i}\int^{\infty}_{-\infty}d\tau\
\bigg(\frac{1}{\tau-x^+}-\frac{1}{\tau-x^-}\bigg)j_k(\tau),
\eea
where $j_k(t)$ is a boundary source
\bea
\lim_{z\rightarrow 0}\Phi_k(x^+, x^-)=j_k(t).
\eea
A boundary term gives the on-shell action or the boundary description
\bea
-\frac{\lambda}{32\pi G_2}\sum_{k=1}^M\int dt\ \Phi_k\partial_z\Phi_k
=-\frac{\lambda}{32\pi^2 G_2}\sum_{k=1}^M\int^{\infty}_{-\infty}d\tau_1d\tau_2\ \frac{j_k(\tau_1)j_k(\tau_2)}{(\tau_1-\tau_2)^2},
\label{onshell}
\eea
in which we use:
\bea
\partial_z\Phi_k&=&
\frac{1}{2\pi}\int^{\infty}_{-\infty}d\tau\
\bigg\lbrack-\partial_{\tau}\bigg(\frac{1}{\tau-x^+}\bigg)-\partial_{\tau}\bigg(\frac{1}{\tau-x^-}\bigg)\bigg\rbrack j_k(\tau),
\nn\\
\lim_{z\rightarrow 0}\partial_z\Phi_k&=&
\frac{1}{\pi}\int^{\infty}_{-\infty}d\tau\ \frac{1}{\tau-t}\partial_{\tau}j_k(\tau).
\eea

\subsection{Deformation of SYK Model}
\noindent
To reproduce the result of the on-shell action \eqref{onshell} in the low-energy limit, we introduce the interaction term to the SYK model \cite{Moitra:2022glw}
\bea
-\frac{1}{\sqrt{M}}\sum_{j=1}^M\sum_{1\le j_1<j_2<j_3<j_4\le N}g_{j_1j_2j_3j_4, j}\psi_{j_1}\psi_{j_2}\psi_{j_3}\psi_{j_4}\phi_j,
\eea
in which the disorder parameter follows the Gaussian distribution \cite{Lau:2023pot,Lau:2025dgd}
\bea
\exp\bigg(-\sum_{j=1}^M\sum_{1\le j_1< j_2< j_3<j_4\le N}g^2_{j_1j_2j_3j_4, j}\frac{N^3}{2g^2}\bigg).
\eea
When integrating out the disorder parameters $g_{j_1j_2j_3j_4, j}$, we obtain four pairs of Majorana fermions \cite{Moitra:2022glw}.
Each pair contributes 1/2 to the denominator.
Therefore, the four pairs of Majorana fermions give the expected denominator as in the on-shell action \eqref{onshell} when the SYK model is dominant \cite{Moitra:2022glw} or considering the following regime:
\bea
G_2\sim\frac{1}{N}\ll 1; \ \phi_b\sim\frac{1}{\beta J}\ll 1; \ \lambda\sim\frac{g^2}{J^3}\ll 1.
\eea
This construction is similar to the AdS/CFT dictionary, in which the scalar field is identified as the boundary source that couples to the dual operator \cite{Anninos:2022qgy}.
Therefore, it is easy to extend this construction to the massive scalar field and also to the fermionic fields \cite{Lau:2023pot,Lau:2025dgd}.
The matter fields can be generated in the large-$N$ limit through the $g$-expansion \cite{Lau:2023pot,Lau:2025dgd}.
JT gravity and its description both agree that the general solution for the dilaton field $\phi_b$ is not affected by the matter coupling \cite{Lau:2023pot,Lau:2025dgd}.
Therefore, the reparametrization symmetry remains, and the bulk time coordinate is dynamical for the boundary description even with the matter coupling \cite{Lau:2023pot,Lau:2025dgd}.
Indeed, it is more general than the conventional AdS/CFT correspondence because the parameter $g$ can be any function of $N$ \cite{Lau:2023pot,Lau:2025dgd}.
The conventional approach using the large-$N$ expansion meets the non-locality issue beyond the cubic matter coupling, and the AdS/CFT dictionary is not practical \cite{Gross:2017hcz}, but we can now deform the distribution of the disorder parameters to generate the quartic matter coupling even in the large-$N$ limit \cite{Lau:2025dgd}
\bea
&&
\exp\Bigg(-\sum_{j=1}^M\sum_{1\le i_1<i_2\le N}g_{i_1i_2, j}^2\frac{N}{2g^2}\Bigg)
\nn\\
&&\times
\exp\Bigg(-\sum_{j=1}^M\sum_{1\le i_1,i_2<k_1, k_2\le N}g_{i_1k_1, j}g_{i_1k_2, j}g_{i_2k_1, j}g_{i_2k_2, j}\frac{3N}{4g^4}\Bigg).
\label{eqn:quarticDistribution}
\eea
Since the bosonic field has an infinite number of degrees of freedom, we need to introduce a cut-off for the simulation \cite{Lau:2023pot,Lau:2025dgd}.
Hence, it is more practical to study the fermionic matter fields for the simulation \cite{Lau:2023pot,Lau:2025dgd}.

\subsection{Random Matrix Theory}
\noindent
The higher-order terms in the matter coupling do not deform the Hamiltonian but deform the disorder distribution \cite{Lau:2025dgd}.
The symmetry of the Hamiltonian is unchanged for each configuration of the disorder parameters.
However, the averaged adjacent gap ratio depends on the disorder-parameter distributions.
However, the deformation of the SYK model for fermionic matter generically extends the RMT classification.
The non-Gaussian effect does not affect the result when $N/M$ is large enough \cite{Lau:2023pot,Lau:2025dgd}.
The patterns of degeneracy and the classifications of random matrix ensembles \cite{Dyson:1962es} can be delineated by the number of Dirac fermions $T_d=N/2+M$ and the charge-conjugation operators when $N/M\gg 1$, as elucidated below:
\begin{itemize}
\item $T_d \bmod 4 = 0, ({\cal P}^2, {\cal R}^2) = (1, 1):$ no degeneracy, GOE;
\item $T_d \bmod 4 = 1, ({\cal P}^2, {\cal R}^2) = (1, -1):$ no degeneracy, GUE;
\item $T_d \bmod 4 = 2, ({\cal P}^2, {\cal R}^2) = (-1, -1):$ twofold degeneracy, GSE;
\item $T_d \bmod 4 = 3, ({\cal P}^2, {\cal R}^2) = (-1, 1):$ no degeneracy, GUE.
\end{itemize}
When $M=0$, the classification is the same as in the case of the SYK model \cite{Cotler:2016fpe}.
It implies that the large-$N$ limit suppresses the non-Gaussian effect, and the Hamiltonian's symmetry universally controls the RMT and degeneracy.

\section{Von Neumann Algebra}
\label{sec:6}
\noindent
We first introduce the vN algebra from its formal mathematical definitions and its application in quantum gravity \cite{Leutheusser:2021qhd,Chandrasekaran:2022cip,Chandrasekaran:2022qmq}.
We then provide examples to clarify the mathematical classification.
The first example is the maximally mixed state, which helps explain the trace operation.
The second and third examples are from the well-defined models of the 2D YM theory and the SYK model \cite{Lau:2023pot,Lau:2025dgd}.

\subsection{Introduction}
\noindent
A vN algebra is a special kind of algebra of bounded operators on a Hilbert space $H$ that is closed under:
\begin{itemize}
\item{addition, multiplication, scalar multiplication,}
\item{adjoint $A \longmapsto A^{\dagger}$.}
\end{itemize}
If $B(H)$ is the algebra of all bounded linear operators on $H$, a vN algebra is a $*$-subalgebra (involutive subalgebra) containing the identity and closed in the weak operator topology (equivalently, in the strong operator topology).
To understand vN algebras, one must distinguish several notions of convergence on $B(H)$:
\begin{itemize}
\item{Norm Topology: $A_n\rightarrow A$ in norm if $||A_n-A||\rightarrow 0$.}
\item{Strong Operator Topology (SOT): $A_n\rightarrow A$ strongly if $||A_n\xi-A\xi||\rightarrow 0$ $\forall\xi\in H$.}
\item{Weak Operator Topology (WOT): $A_n\rightarrow A$ weakly if $\langle\eta, A_n\xi\rangle\rightarrow\langle\eta, A\xi\rangle$ $\forall\xi, \eta\in H$.
This is weaker than strong convergence. }
\end{itemize}

\noindent
The most famous equivalent formulation is the double commutant theorem.
Given a subset $S\subset B(H)$, its commutant is
\bea
S^{\prime}=\{T\in B(H): \ TQ=QT \ \forall Q\in S\}.
\eea
Therefore, $S^{\prime}$ consists of all bounded operators commuting with everything in $S$.
We then define the double commutant
\bea
S^{\prime\prime}=(S^{\prime})^{\prime}.
\eea
A basic fact: $S^{\prime\prime}$ is the smallest vN algebra naturally generated by $S$.
If $A\in B(H)$, the following are equivalent
\begin{itemize}
\item{$A$ is weakly closed,}
\item{$A$ is strongly closed,}
\item{$A=A^{\prime\prime}$.}
\end{itemize}

\noindent
Physically, vN algebras formalize the algebra of observables of a quantum system.
They are especially important when:
\begin{itemize}
\item{the Hilbert space is infinite-dimensional,}
\item{one studies thermodynamic limits,}
\item{one wants a local algebra of observables in QFT,}
\item{one wants a clean framework for states, measurements, entropy, and operator algebras.}
\end{itemize}
They are more intimately connected to operator-theoretic and measure-theoretic frameworks.

\subsubsection{Projections}
\noindent
A projection is an operator $P$ such that
\bea
P=P^2=P^*.
\eea
Projections represent yes/no measurements.
We have the following projections
\begin{itemize}
\item{minimal projection $P_m$: the projection $P\le P_m\neq 0$$\Rightarrow$ $P=P_m$ or $P=0$;}
\item{finite projection $P_f$: a projection is finite if it is not equivalent to a proper subprojection of itself.
Otherwise, it is infinite.}
\end{itemize}

\subsubsection{States}
A state on a vN algebra ${\cal M}$ is a linear functional such that
\begin{itemize}
\item{$\omega(x^*x)\ge0$,}
\item{$\omega(I)=1$.}
\end{itemize}
For ${\cal M}=B(H)$, normal states are exactly density-matrix states:
\bea
\omega(A)=\mathrm{Tr}(\rho A); \ \rho\ge 0; \ \mathrm{Tr}\rho=1.
\eea
If we consider the following density matrix
\bea
\rho=\frac{1}{N}\sum_{j=1}^N|j\rangle\langle j|
\eea
with the infinite $N$ limit $N\rightarrow\infty$,  the density matrix is not normalizable.
Hence, the state is not normal.
QFT usually considers the finite but large $N$ to treat this non-normal state, and takes the infinite $N$ limit just for extracting the finite observables.
Indeed, normal states generalize density matrices to arbitrary vN algebras.

\subsubsection{Center and Factor}
\noindent
The center of ${\cal M}$ is
\bea
Z({\cal M})={\cal M}\bigcap{\cal M}^{\prime}.
\eea
These are the elements that commute with everything in ${\cal M}$.
A vN algebra is designated as a factor if its center comprises solely those operators that are proportional to the identity operator.
In general, vN algebras can frequently be expressed as direct integrals of factors:
\bea
{\cal M}\cong\int^{\oplus}d\mu(x)\ {\cal M}_x,
\eea
where each ${\cal M}_x$ is classified as a factor.
Therefore, factors are the irreducible building blocks of all vN algebras.

\subsubsection{Classification of Factors: Types I, \II, \III}
\noindent
For a factor, there is no non-trivial center, so the only way to distinguish the structure is through:
\begin{itemize}
\item{projections;}
\item{equivalence between projections;}
\item{existence (or not) of traces,}
\end{itemize}
where a trace on ${\cal M}$ is a positive linear function $\tau$ satisfying
\bea
\tau(xy)=\tau(yx).
\eea
In QFT, we allow two operators to fail to commute under the trace operation, as in the commutation relation between the momentum and position operators in scalar field theories.
However, two operators must commute under the trace operation in the context of vN algebra.
Therefore, the classification is fundamentally about how projections behave inside the algebra.
There are three fundamentally different behaviors:
\begin{itemize}
\item Type I (Discrete):
\begin{itemize}
\item There exist minimal projections;
\item Type I factor $\cong$ $B(H)$;
\item The usual matrix trace works in type I$_n$, which implies that the bounded operators on a Hilbert space of finite dimension $n$ are factors of type I$_n$.
\end{itemize}
\item Type \II (Continuous but Finite):
\begin{itemize}
\item No minimal projections (or normal states), but there exist finite projections and a trace;
\item In type \II$_1$, there is a normalized trace with values in $[0, 1]$ on projections.
\end{itemize}
\item Type \III (Purely Infinite/No Trace):
\begin{itemize}
\item No trace, pure states, or density matrices exist at all, and every non-zero projection is infinite.
\end{itemize}
\end{itemize}
The intuition for each classification is:
\begin{itemize}
\item{Type I: quantum mechanics of particles with factorization of Hilbert space;}
\item{Type \II: continuous probability-like systems rather than discrete, but still trace-like counting exists;}
\item{Type \III: quantum fields with infinite short-distance entanglement that lead to everything being "infinitely large," losing finite or normalized measures or density matrix internal to the algebra.}
\end{itemize}
We can think that the classification is the same as:
\bea
&&
\mathrm{vN\ algebra}\longrightarrow\mathrm{decompose\ into\ factors}
\nn\\
&&
\longrightarrow\mathrm{classify\ each\ factor\ by\ projection\ structure}.
\eea

\subsubsection{Quantum Field Theory and Holography}
\noindent
In quantum mechanics, or in a bipartite system
\bea
H=H_A\otimes H_B,
\eea
the algebra of observables of subsystem $A$ is
\bea
{\cal A}=B(H_A)\otimes I_B.
\eea
This represents a type I factor.
Then:
\begin{itemize}
\item {the reduced density matrix exists,}
\item{the commutant is
\bea
{\cal A}^{\prime}=I_A\otimes B(H_B),
\eea
}
\item{the Hilbert space factorizes cleanly.}
\end{itemize}
This picture works well for:
\begin{itemize}
\item{finite-dimensional systems;}
\item{lattice systems with finitely many sites;}
\item{quantum mechanics of finitely many particles.}
\end{itemize}

\noindent
In local QFT, to each spacetime region $O$, one assigns a vN algebra
\bea
{\cal A}(O)\in B(H),
\eea
interpreted as the algebra generated by all observables localized in $O$.
These satisfy properties such as:
\begin{itemize}
\item{Isotony: If $O_1\in O_2$, we have ${\cal A}(O_1)\in {\cal A}(O_2)$.}
\item{Locality: If $O_1$ and $O_2$ are spacelike separated, we have $\lbrack{\cal A}(O_1), {\cal A}(O_2)\rbrack=0$.}
\item{Covariance: Symmetries act geometrically on regions and correspondingly on the algebras.}
\end{itemize}
The natural object in the local QFT is not first the tensor factor $H_A$, but the local operator algebra ${\cal A}(O)$.
This shift is crucial.
In continuum local QFT, one might hope that for a spatial region $O$ to have the Hilbert space decomposition,
\bea
H=H_O\otimes H_{O^c},
\eea
with
\bea
{\cal A}(O)=B(H_O)\otimes I; \ {\cal A}(O^c)=I\otimes B(H_{O^c}),
\eea
but this fails in general.
The intuitive reason is
\begin{itemize}
\item{field degrees of freedom live at arbitrarily short distances,}
\item{near the boundary $\partial O$, modes inside and outside are infinitely entangled,}
\item{this ultraviolet entanglement is not a small correction; it changes the operator-algebraic type.}
\end{itemize}
The local algebra is too large and too singular to be type I, but too local to be the whole $B(H)$.
The result is a typical type \III.
This precisely elucidates why entanglement entropy in quantum field theory is subtle.
\\

\noindent
A type I factor has minimal projections, like rank one projectors $|\psi\rangle\langle\psi|$, but in local QFT algebras, they do not behave like that.
There are no such minimal local projections.
Physically, this matches the idea that one cannot isolate the smallest local degrees of freedom at a point in a continuum field theory.
A typical \III factor goes much further:
\begin{itemize}
\item{it has no non-zero finite projections;}
\item{it admits no normalizable trace;}
\item{hence no intrinsic density matrix formalism analogous to finite subsystems.}
\end{itemize}
Hence, local QFT algebras are not merely "infinite-dimensional type I"; they are a different species.
Therefore, vN algebra theory explains mathematically why continuum local QFT does not factorize the way lattice systems do.
\\

\noindent
In type I quantum mechanics, a normal state on $B(H_A)$ can be articulated as
\bea
\omega(a)=\mathrm{Tr}(\rho_A a).
\eea
This formula depends on the existence of a trace and a reduced density matrix $\rho_A$.
For type \III algebras, there is no suitable trace of this sort.
Consequently, the state confined to a local algebra ${\cal A}(O)$ retains substantial significance as a positive linear functional; however, it is not ubiquitously articulated as
\bea
\omega_O(a)=\mathrm{Tr}(\rho_Oa)
\eea
for some reduced density matrix $\rho_O$ intrinsic to the algebra.
This encapsulates the algebraic interpretation of "no reduced density matrix" within the continuum framework.
One can still use regulator-dependent density matrices after imposing a cutoff.
However, they are not fundamental continuum objects, and the algebra is effectively shifted from type \III to type I.
\\

\noindent
For type \II algebras, we lose the normal state, but we can still define $\omega_O(a)$ by the trace operation.
Even though we are restricted to non-normalizable states, the divergence can be effectively mitigated through a suitable trace operation \cite{Chandrasekaran:2022cip}.
This scenario resembles the ultraviolet divergence in the entanglement entropy, which can be resolved by integrating out the matter and pure gravity sectors \cite{Chandrasekaran:2022cip}.
Hence, the QFT on the curved background provides a better understanding of the computation of the entanglement entropy compared to pure QFT or pure gravity \cite{Chandrasekaran:2022cip}.
The semi-classical theory belongs to the type \II algebra \cite{Chandrasekaran:2022cip}.
Because the type I algebra allows all bounded operators, the algebra of quantum gravity must correspond to it \cite{Chandrasekaran:2022cip}.
Hence, we can always start with type I algebra and observe how type \II or type \III algebras emerge from it.
For the SYK model, the number of Majorana fermions $N$ controls the strength of the gravitational constant when $N$ is large enough \cite{Polchinski:2016xgd,Maldacena:2016hyu}.
Hence, we can start from the finite-$N$ case and then observe the emergent gravity from the large-$N$ expansion \cite{Chandrasekaran:2022qmq}.
Because the different kinds of the vN algebra have different properties, we can distinguish the algebras by computing observables.
This helps clarify the relationship between emergent gravity and the SYK model because the vN algebra provides a general classification of operators.

\subsection{Examples}
\noindent
We introduce examples to explain the abstract definitions.
The maximally mixed state can serve as a simple demonstration of the trace operation.
We consider the 2D YM theory to illustrate how to treat operator algebras in a specific theory.
For the SYK model, we explain how the choice of operator and state setup affects the results.

\subsubsection{Maximally Mixed State}
\noindent
Given a Hamiltonian $H_N$ with $N$ qubits, the thermal state is
\bea
\omega_{N, \beta}(A)=\frac{\mathrm{Tr}_H(e^{-\beta H_N}A)}{Z_N(\beta)},
\eea
where
\bea
Z_N(\beta)=\mathrm{Tr}_H(e^{-\beta H_N}).
\eea
The $\mathrm{Tr}_H$ is the Hilbert-space trace, distinguished from the trace of the vN algebra $\tau$.
Since $\exp(-\beta H_N)=I$ at $\beta=0$, we get
\bea
\omega_{N, 0}(A)=\frac{1}{2^N}\mathrm{Tr}_H(A).
\eea
This density matrix represents the maximally mixed state of an $N$-qubit system.
\bea
\rho_{N, 0}=\frac{1}{2^N}I.
\eea
Thereofre, for finite $N$, we have
\bea
\beta=0\Longleftrightarrow \ \mathrm{maximally\ mixed\ state}\Longleftrightarrow \ \mathrm{normalized\ trace}.
\eea
At $\beta\neq 0$, the state becomes
\bea
\rho_{N, \beta}=\frac{e^{-\beta H_N}}{Z_N(\beta)},
\eea
which is not proportional to the identity unless $H_N$ is trivial.
Therefore, it is not maximally mixed.
\\

\noindent
Now, take the local maximally mixed states
\bea
\omega_{\Lambda, 0}(A)=\frac{1}{2^{|\Lambda|}}\mathrm{Tr}_H(A).
\eea
These are compatible as $\Lambda$ grows, so they define a global state $\tau$ at infinite temperature and volume.
It satisfies
\bea
\tau(AB)=\tau(BA)
\eea
for all local observables.
Hence, $\tau$ is a trace.
For finite $N$, the maximally mixed state is proportional to the identity.
For infinite qubits, there is no density matrix
\bea
\rho\propto I
\eea
on a separable Hilbert space that represents a normal state on the full algebra, because the identity would have infinite trace, but the algebraic content of "maximally mixed" survives:
\begin{itemize}
\item{it is the trace,}
\item{it is invariant under local unitaries,}
\item{it gives no preferred basis.}
\end{itemize}
Hence, in the infinite system, the maximally mixed state does not mean the "trace-class density matrix proportional to identity".
The resulting algebra is type \II$_1$, not type I, because the $\beta=0$ state preserves a normalized trace in the infinite-volume limit.
\\

\noindent
In finite matrices, we have
\bea
\tau_N(A)=\frac{1}{2^N}\mathrm{Tr}_H(A).
\eea
The limit retains this trace property.
The identity still has finite total weight:
\bea
\tau(I)=1.
\eea
The maximally mixed state reflects the existence of a normalized trace, which is the defining hallmark of a finite factor; its breakdown in the large-$N$ limit signals the transition toward the operator-algebraic structure underlying emergent gravity.
Since the algebra is infinite-dimensional and has no minimal projections, the factor is type \II$_1$, not type I$_n$.
Since it has a trace, it is not type \III.
Hence, we conclude that:
\begin{itemize}
\item{infinite-dimensional+trace+finite factor$\Rightarrow$\II$_1$.}
\item{not $I_{\infty}$, because $B(H)$ has no normalized trace,}
\item{not \III, because the trace exists.}
\end{itemize}
A concrete model is an infinite spin chain.
For each site $j$, let the one-site state be
\bea
\tau_j(a)=\frac{1}{2}\mathrm{Tr}(a),
\eea
where $a$ is a $ 2\times 2$ matrix.
Then the infinite product state
\bea
\tau=\bigotimes_{j\in\mathbb{Z}}\tau_j
\eea
is the infinite-temperature state.
For a local observable
\bea
A=a_{j_1}\otimes a_{j_2}\otimes\cdots\otimes a_{j_n},
\eea
one has
\bea
\tau(A)\prod_{k=1}^n\frac{1}{2}\mathrm{Tr}_H(a_{j_k}).
\eea
Hence, this is the cleanest concrete realization of "maximally mixed infinite qubits."
Now, let us move to $\beta\neq 0$.
For a finite region $\Lambda$, define the thermal state
\bea
\omega_{\Lambda, \beta}(A)=\frac{\mathrm{Tr}_H\big(e^{-\beta H_{\Lambda}}A\big)}{Z_{\Lambda}(\beta)}.
\eea
At finite $\Lambda$, this is still a density matrix state on a matrix algebra, so everything is type I there.
However, the real question is what happens after taking the thermodynamic limit $\Lambda\rightarrow\infty$.
The observable algebra can become type \III, where no normalized trace or density-matrix description exists.
At $\beta=0$, we have
\bea
\omega_{\Lambda, 0}(AB)=\omega_{\Lambda, 0}(BA)
\eea
because the state is the normalized trace.
At $\beta\neq 0$, we have
\bea
\omega_{\Lambda, \beta}(A)=\frac{\mathrm{Tr}_H\big(e^{-\beta H_{\Lambda}}A\big)}{Z_{\Lambda}(\beta)},
\eea
and unless $H_{\Lambda}$ is proportional to identity,
\bea
\omega_{\Lambda, \beta}(AB)\neq\omega_{\Lambda, \beta}(BA).
\eea
Hence, the state is not a trace state.
\\

\noindent
Now the main point: For many infinite quantum spin systems, the non-zero temperature state yields a vN algebra that is not finite and often is type \III.
The intuition is:
\begin{itemize}
\item{the infinite system has infinitely many degrees of freedom,}
\item{thermal weight is distributed across infinitely many sites,}
\item{there is no normalized trace on the resulting equilibrium algebra,}
\item{the thermodynamic-limit equilibrium state is not represented by a trace-class density matrix.}
\end{itemize}
The algebraic structure shifts:
\bea
\beta=0\Longrightarrow \mathrm{\II}_1,
\eea
while
\bea
\beta\neq0\Longrightarrow\mathrm{often\ type\ \III}.
\eea
The precise type of the non-zero thermal equilibrium state in an infinite system depends on:
\begin{itemize}
\item{the dynamics,}
\item{the interaction,}
\item{the dimension,}
\item{whether there are phase transitions,}
\item{whether one is at a critical point.}
\end{itemize}

\noindent
A very instructive example is independent qubits with a one-site Hamiltonian
\bea
h=\frac{\epsilon}{2}\sigma_z.
\eea
The one-site thermal state is
\bea
\omega_{\beta}^{(j)}(a)=\frac{\mathrm{Tr}_H\big(e^{-\beta h}a\big)}{\mathrm{Tr}_H\big(e^{-\beta h}\big)}.
\eea
Its density matrix is
\bea
\rho_{\beta}^{(j)}=\frac{e^{-\beta h}}{\mathrm{Tr}_H\big(e^{-\beta h}\big)}=
\begin{pmatrix}
p&0
\\
0&1-p
\end{pmatrix},
\eea
where $p\neq 1/2$ for $\beta\neq 0$.
The infinite product state is
\bea
\omega_{\beta}=\bigotimes_{j\in\mathbb{Z}}\omega_{\beta}^{(j)}.
\eea
At $\beta=0, p=1/2$, the one-site state is a trace, and the infinite product yields type \II$_1$.
At $\beta\neq 0, p\neq1/2$, so the one-site state is not a trace
\bea
\mathrm{Tr}_H(\rho_{\beta}^{(j)}ab)\neq\mathrm{Tr}_H(\rho_{\beta}^{(j)}ba).
\eea
Then the infinite tensor product vN algebra is no longer \II$_1$; for generic parameters, it is of type \III.
Hence, this example clearly shows the transition.
Even though each finite subsystem has
\bea
\rho_{\Lambda, \beta}=\frac{e^{-\beta H_{\Lambda}}}{Z_{\Lambda}},
\eea
the infinite system generically does not admit a single global trace-class operator $\rho_{\beta}$ with
\bea
\omega_{\beta}(A)=\mathrm{Tr}(\rho_{\beta} A)
\eea
on a global $B(H)$.
This is because:
\begin{itemize}
\item{the thermodynamic (infinite-volume) limit destroys the type I trace-class framework,}
\item{the partition function is no longer finite, so the thermal state cannot be normalized as a density matrix.}
\end{itemize}
Hence, again, the correct description is algebraic, not density-matrix-based.
There is a clean intuition for the \II$_1$$\rightarrow$\III transition here.
At $\beta=0$, local states behave as maximally mixed.
There is no energetic bias.
Therefore, the state is a trace.
Hence, finite-type behavior survives, giving \II$_1$.
At $\beta\neq 0$, the state now distinguishes between energy levels.
The thermodynamic limit removes the trace-class structure; the trace no longer exists.
In the infinite system, that typically forces the equilibrium representation into type \III.
\\

\noindent
The infinite spin systems are related but not identical to local QFT:
\begin{itemize}
\item{For infinite spin systems, $\beta=0$ can give \II$_1$, and $\beta\neq0$ can give type \III.}
\item{For continuum local QFT, even the vacuum local algebra is already typically type \III because of ultraviolet entanglement and locality.}
\end{itemize}
Therefore, in spin systems, type \III can emerge from the thermodynamic limit.
In continuum local QFT, type \III is already built into the local structure even at zero temperature.
These are two different roads to type \III.
\\

\noindent
When we lose the trace of the vN algebra, the observables can be different depending on the Hilbert space representations.
A presentation is
\bea
\pi: {\cal A}\rightarrow B(H),
\eea
where ${\cal A}$ is the algebra of observables.
Two representations $\pi_1, \pi_2$ are:
\begin{itemize}
\item{equivalent if related by a unitary;}
\item{inequivalent otherwise.}
\end{itemize}
A physically meaningful quantity has the same value in all representations, i.e., it depends only on ${\cal A}$, not on $H$.
The Hilbert-space trace $\mathrm{Tr}_H\big(\pi(A)\big)$ depends explicitly on the choice of the Hilbert space representation.
Therefore, it is not intrinsic to ${\cal A}$.
Let us do a very explicit example.
\bea
\pi_1(A)=A; \
\pi_2(A)=
\begin{pmatrix}
A&0
\\
0&A
\end{pmatrix},
\eea
where $A$ is a 2 by 2 matrix.
We now compute the Hilbert-space trace:
\bea
\mathrm{Tr}_{H_1}\big(\pi_1(A)\big)=\mathrm{Tr}(A); \
\mathrm{Tr}_{H_2}\big(\pi_2(A)\big)=2\mathrm{Tr}(A).
\eea
Hence, the Hilbert-space trace depends on the representation because the Hilbert space dimension changes.
The trace of the vN algebra, such as
\bea
\tau(A)=\frac{1}{2}\mathrm{Tr}(A),
\eea
does not change because $\tau(I)=1$ is defined directly on the algebra and does not depend on the Hilbert space.
Therefore, $\tau$ represents the expectation value intrinsic to observables, not the same as $\mathrm{Tr}_H$.
The $\mathrm{Tr}_H$ counts states in $B(H)$, but $\tau$ measures ${\cal A}$.
Hence, in type \II$_1$, $\tau(I)=1$ replaces $\mathrm{Tr}_H(I)=\mathrm{dimension}$.

\subsubsection{2D YM Theory}
\noindent
The Lagrangian for 2D YM theory with the Euclidean signature is
\bea
{\cal L}_E=\frac{1}{4g_{\mathrm{YM}}^2}g^{\mu_1\mu_2}g^{\nu_1\nu_2}F_{\mu_1\nu_1}^aF_{\mu_1\nu_2}^a,
\eea
where $g_{\mu\nu}$ is the background metric, $a$ is Lie algebra index, and $\mu, \nu$ are the spacetime indices.
The non-Abelian field strength is
\bea
F_{\mu\nu}^a\equiv \partial_{\mu}A_{\nu}^a-\partial_{\nu}A_{\mu}^a+f^{abc}A_{\mu}^bA_{\nu}^c,
\eea
where $f_{abc}$ is the structure constant with the anti-symmetric indices.
For the fundamental representation of SU($N$), we have the following normalization and the trace formula:
\bea
\mathrm{Tr}(T^aT^b)=\frac{1}{2}\delta^{ab}; \
\sum_a (T^a)_{jk}(T^a)_{lm}=\frac{1}{2}\bigg(\delta_{jm}\delta_{lk}-\frac{1}{N}\delta_{jk}\delta_{lm}\bigg),
\eea
where $T^a$ is the generator of the SU($N$).
The algebra for the generators is given by:
\bea
\lbrack T^a, T^b\rbrack= f^{abc}T^c; \
\{ T^a, T^b\}=\frac{1}{N}\delta^{ab}+d^{abc}T^c,
\eea
where $d^{abc}$ is a structure constant with symmetric indices.
\\

\noindent
We obtain the following identity
\bea
-\bigg\langle\frac{1}{g_{\mathrm{YM}}^2}\big(\nabla_{\mu}F^{\mu\nu}(x)\big)^aQ[A]\bigg\rangle
=\bigg\langle\frac{\delta}{\delta A_{\nu}^a(x)}Q[A]\bigg\rangle,
\eea
where
\bea
\nabla_{\mu}\equiv\partial_{\mu}+A_{\mu},
\eea
by requiring
\bea
0=\int{\cal D}A\ \frac{\delta}{\delta A_{\nu^a}}\bigg(e^{-S_E}Q[A]\bigg).
\eea
Now we substitute the Wilson line for $Q[A]$ and obtain
\bea
-\bigg\langle\frac{1}{g_{\mathrm{YM}}^2}\mathrm{Tr}\big(\nabla_{\mu}F^{\mu\nu}(x)\big) e^{i\oint_Cd\xi^{\mu} A_{\mu}}\bigg\rangle
=\bigg\langle\mathrm{Tr}\frac{\delta}{\delta A_{\nu}(x)}e^{i\oint_Cd\xi^{\mu} A_{\mu}}\bigg\rangle.
\eea
We can calculate the variation from the SU($N$) case
\bea
\frac{\delta A_{\mu}^{jk}(y)}{\delta A_{\nu}^{mn}(x)}
=\frac{1}{2}\delta_{\mu\nu}\delta^2(x-y)
\bigg(\delta^{jn}\delta^{mk}-\frac{1}{N}\delta^{jk}\delta^{mn}\bigg),
\eea
where
\bea
A_{\mu}^{jk}(x)\equiv A_{\mu}^a(x)(T^a)^{jk}.
\eea
This formula is due to the fact that
\bea
\frac{\delta A_{\mu}^a(y)}{\delta A_{\nu}^b(x)}
=\delta_{\mu\nu}\delta^2(x-y)\delta^{ab}.
\eea
Hence, we obtain
\bea
&&
\mathrm{Tr}\frac{\delta}{\delta A_{\nu}(x)}e^{i\oint_Cd\xi^{\mu}A_{\mu}}
\nn\\
&=&\frac{i}{2}\oint_C dy_{\nu}\ \delta^2(x-y)
\bigg\lbrack\frac{1}{N}\mathrm{Tr}\bigg(e^{i\int_{C_{yx}}d\xi^{\mu}A_{\mu}}\bigg)
\frac{1}{N}\mathrm{Tr}\bigg(e^{i\int_{C_{xy}}d\xi^{\mu}A_{\mu}}\bigg)
-\frac{1}{N^3}\mathrm{Tr}\bigg(e^{i\int_{C}d\xi^{\mu}A_{\mu}}\bigg)
\bigg\rbrack.
\nn\\
\eea
Finally, we get
\bea
&&
-\bigg\langle\frac{1}{g_{\mathrm{YM}}^2}\mathrm{Tr}\big(\nabla_{\mu}F^{\mu\nu}(x)\big) e^{i\oint_Cd\xi^{\mu} A_{\mu}}\bigg\rangle
\nn\\
&=&\frac{i}{2}\oint_Cdy_{\nu}\ \delta^2(x-y)\bigg(\langle\Phi(C_{yx})\Phi(C_{xy})\rangle-\frac{1}{N^2}\langle\Phi(C)\rangle\bigg),
\eea
where
\bea
\Phi(C)\equiv \frac{1}{N}\mathrm{Tr}\bigg(e^{i\oint_Cd\xi^{\mu} A_{\mu}}\bigg).
\eea
Therefore, we can define a normalized trace in the strict large-$N$ limit
\bea
\mathrm{TR}({\cal O})\equiv\bigg\langle\frac{1}{N}\mathrm{Tr}({\cal O})\bigg\rangle
\eea
because we have the large-$N$ factorization
\bea
&&
-\bigg\langle\frac{1}{g_{\mathrm{YM}}^2}\mathrm{Tr}\big(\nabla_{\mu}F^{\mu\nu}(x)\big) e^{i\oint_Cd\xi^{\mu} A_{\mu}}\bigg\rangle
\nn\\
&=&\frac{i}{2}\oint_Cdy_{\nu}\ \delta^2(x-y)\langle\Phi(C_{yx})\rangle\langle\Phi(C_{xy})\rangle+{\cal O}\bigg(\frac{1}{N^2}\bigg).
\eea
Hence, it is a type \II$_1$ algebra.

\subsubsection{Deformation of SYK Model}
\noindent
When considering the deformation of the SYK model for the fermionic matter, the physical degrees of freedom are finite for any finite number of $N$ Majorana fermions and $M$ Dirac fermions \cite{Lau:2023pot,Lau:2025dgd}.
Hence, we have a factorization of the Hilbert space into Majorana and Dirac fermion sectors \cite{Lau:2023pot,Lau:2025dgd}.
According to the low-energy theory, the entanglement between the Majorana and Dirac fermions in the strict large-$N$ limit corresponds to the entanglement between the graviton and matter fields \cite{Lau:2023pot,Lau:2025dgd}.
The late-time behavior of deformed SYK models indicates that the emergent gravitational degrees of freedom—identified with the graviton—become maximally entangled with the matter fields \cite{Lau:2023pot,Lau:2025dgd}.
It seems to contradict the idea that classical gravity cannot have quantum entanglement.
However, a quantum system's classical limit means the observable satisfies the classical equations, while the state remains quantum.
Hence, the bulk theory's classical limit corresponds to a large, entangled system with suppressed fluctuations.
It is similar to the relation between quantum and classical chaos.
A quantum state that satisfies a linear equation is expected to exhibit nonchaotic dynamics.
However, its classical limit can be chaotic because the chaotic property lies in the observable's dynamical behavior, not in the quantum state.
\\

\noindent
The result of the maximally entangled state for $N=2M\rightarrow\infty$ at late time implies equal degrees of freedom between the Majorana and Dirac fermions, which shows the type \II$_1$ algebra \cite{Lau:2023pot,Lau:2025dgd}.
It realizes the type I-to-type \II algebraic transition \cite{Lau:2023pot,Lau:2025dgd}.
Because the type \II loses the normal state, some information is lost.
Hence, the algebraic transition helps explain the information loss associated with the mechanism of emergent spacetime \cite{Lau:2023pot,Lau:2025dgd}.
For other values of $N$, it is ambiguous which algebra to identify with, depending on the setup.
As we vary $N$ and $M$, the Hilbert space changes.
Therefore, the operator content also changes.
If we introduce all operators for all possible $N$ and $M$, this model's late-time result for $N\neq 2M$ is a type \II$_1$ algebra \cite{Lau:2023pot,Lau:2025dgd}.
However, if we specifically consider operators for the particular values of $M$ in the strict large-$N$ limit, the algebra need not be of type \II; it can be type I.
Hence, the result depends on the set-up, which in turn depends on the motivation and goal.

\section{Outlook and Future Directions}
\label{sec:7}
\noindent
In this review, we have examined various aspects of quantum information in the context of the Sachdev–Ye–Kitaev (SYK) model.
One of the most compelling open problems in this direction is to understand the emergence of spacetime using quantum information–theoretic tools.
Establishing such a connection would provide important insights into the microscopic mechanism underlying emergent spacetime in quantum gravity.
Recent developments in quantum information have introduced the perspectives of quantum chaos and entanglement into the study of the holographic principle.
In particular, the chaos bound on the Lyapunov exponent \cite{Larkin:1969} for simple operators in conformal field theories provides a direct link to semiclassical Einstein gravity in asymptotically Anti–de Sitter (AdS) spacetimes \cite{Maldacena:2015waa,Narovlansky:2025tpb}.
\\

\noindent
From an operator-algebraic viewpoint, von Neumann (vN) algebras provide a natural classification of boundary quantum systems and their corresponding bulk gravitational descriptions in terms of their entanglement structure \cite{Leutheusser:2021qhd,Chandrasekaran:2022cip}.
Together, these approaches establish a concrete connection between quantum chaos \cite{Berry:1977zz,Berry:1977wpp}, entanglement, and emergent spacetime. Understanding this connection is also crucial for addressing the black hole information paradox.
While quantum gravity is expected to preserve information, semiclassical gravity appears to lead to information loss.
This tension suggests that spacetime itself may not be fundamental but instead emerges from the low-energy dynamics of an underlying quantum theory.
\\

\noindent
The SYK model \cite{Polchinski:2016xgd,Maldacena:2016hyu} provides a particularly tractable holographic system, with analytical control over various quantum information measures.
In the large-$N$, low-energy limit, the SYK model reproduces the effective action of Jackiw–Teitelboim (JT) gravity, establishing a duality between a quantum mechanical system and two-dimensional gravity \cite{Polchinski:2016xgd,Maldacena:2016hyu}.
Moreover, nonperturbative features of the model can be explored numerically, even in regimes where analytic methods are unavailable.
The SYK spectrum exhibits random-matrix-theory (RMT) statistics, offering nontrivial insights into the relationship between quantum chaos and semiclassical gravity \cite{Garcia-Garcia:2016mno,Garcia-Garcia:2017pzl,Dyer:2016pou,Cotler:2016fpe}.
\\

\noindent
Because the SYK model has finitely many degrees of freedom at any finite $N$, its operator algebra is initially of type I, allowing for a density-matrix description and a well-defined trace \cite{Chandrasekaran:2022qmq}.
This provides a controlled setting to study the transition to more general von Neumann algebra structures—such as type \II or type \III—in the large-$N$ limit, which is expected to play a key role in the emergence of spacetime \cite{Chandrasekaran:2022qmq}.
However, the SYK model also exhibits nonlocal features when the AdS/CFT dictionary is extended beyond leading orders, making a complete holographic reconstruction challenging.
\\

\noindent
To address these limitations, one may consider deformations of the SYK model that introduce additional matter degrees of freedom through independent coupling parameters in the strict large-$N$ limit \cite{Lau:2023pot,Lau:2025dgd}.
In this regime, the resulting theory admits a local effective description that includes interacting matter sectors.
When restricted to fermionic matter fields, each finite-$N$ realization still contains only finitely many degrees of freedom, preserving analytical and numerical tractability.
These deformed models therefore provide a promising framework for systematically exploring the interplay between quantum chaos, entanglement, and operator-algebraic structures in holography.
\\

\noindent
Deformations of the SYK model provide a mechanism for generating additional matter fields in the strict large-$N$ limit.
This construction extends the standard SYK–JT correspondence and continues to reproduce the effective action of Jackiw–Teitelboim (JT) gravity at low energies, now coupled to matter sectors.
Despite this promising structure, a detailed analysis of the deformed model's chaotic properties remains largely unexplored.
In particular, the spectrum of the kernel governing four-point functions has not been systematically studied, and the associated Lyapunov exponent has not yet been computed.
Such an analysis is crucial for establishing whether the deformed SYK model fully complies with the holographic dictionary.
In analogy with the original SYK model, one expects that the Lyapunov exponent for simple operators saturates the universal chaos bound, reflecting the presence of semiclassical gravitational dynamics in the bulk \cite{Maldacena:2015waa,Narovlansky:2025tpb}.
This expectation is motivated by the fact that the dual gravitational description corresponds to JT gravity \cite{Teitelboim:1983ux,Jackiw:1984je} coupled to matter fields, where maximal chaos is typically associated with near-horizon dynamics.
Verifying this behavior would provide an important consistency check of the extended SYK–gravity correspondence and clarify the role of matter couplings in holographic quantum chaos.
\\

\noindent
The fermionic matter deformation of the SYK model preserves the universal features of random matrix theory (RMT) \cite{Lau:2023pot,Lau:2025dgd}.
In particular, both Gaussian and non-Gaussian distributions of the random couplings yield essentially identical spectral statistics.
This indicates that the classification into RMT universality classes is governed primarily by the symmetry of the Hamiltonian, rather than the detailed form of the disorder distribution.
This result is somewhat surprising, as one might expect the coupling distribution to influence quantities such as the averaged adjacent gap ratio.
However, any such dependence appears to be strongly suppressed in practice.
Nevertheless, from a theoretical perspective, it is natural to expect that the choice of disorder distribution should, in general, affect the effective theory and potentially modify the associated RMT classification.
Since the low-energy dynamics—and hence the emergent bulk gravitational description—depend on the statistical properties of the couplings, these effects may carry important implications for holography \cite{Lau:2023pot,Lau:2025dgd}.
In particular, the disorder parameter effectively mixes configurations with varying degrees of integrability \cite{Lau:2020qnl,Ozaki:2025mma}, suggesting that interpreting integrable versus non-integrable behavior solely from spectral statistics may require refinement in disordered systems.
\\

\noindent
To clarify these issues, it is important to investigate a broader class of models with more general disorder structures.
In particular, incorporating higher-order deformations in non-Gaussian ensembles and exploring alternative distributions could provide new analytical control over the corresponding bulk gravitational theories.
Such studies may reveal how microscopic disorder data is encoded in the emergent geometry and further illuminate the role of randomness in holographic duality.
\\

\noindent
The SYK model provides a concrete realization of holography in an AdS$_2$ background through its large-$N$, low-energy limit.
In contrast, extending this correspondence to de Sitter (dS) backgrounds remains much less well understood.
Existing discussions are largely based on symmetry arguments for the matter sector.
In contrast, a complete description of the pure gravitational sector in dS remains lacking.
\\

\noindent
A particularly intriguing direction arises from the observation that deformations of the SYK model exhibit a type II$_1$ von Neumann algebra structure at late times \cite{Lau:2023pot,Lau:2025dgd}.
This suggests a possible correspondence with dS black holes, where the system is expected to approach thermal equilibrium \cite{Leutheusser:2021qhd,Chandrasekaran:2022cip}.
In this picture, the late-time state corresponds to a single fixed point of the dynamics, reflecting equilibration within a finite trace framework.
This perspective raises an important question: whether introducing matter couplings can drive the system toward a different fixed point, thereby modifying the late-time algebraic structure and its gravitational interpretation.
Addressing this issue requires a detailed analysis of dynamical observables, such as two-point correlation functions.
Such studies may help uncover the dynamical mechanism governing the transition from early-time AdS$_2$ behavior to a late-time dS regime, and clarify how matter interactions influence the emergent spacetime geometry \cite{Lau:2023pot,Lau:2025dgd}.

\section*{Acknowledgments}
\noindent
JM would like to acknowledge support from the ``Quantum Technologies for Sustainable Development''  grant from the National Institute for Theoretical and Computational Sciences of South Africa (NITHECS).
MT acknowledges the Grants-in-Aid from MEXT of Japan (Grants No. JP21H05185, and JP25K00925) and JST CREST (Grant No. JPMJCR24I2). 
CTM thanks Nan-Peng Ma for his encouragement.



\end{document}